\documentclass[%
 preprint, %linenumbers,
 amsmath,amssymb,
 aps, prfluids,
]{revtex4-2}

\usepackage{graphicx}% Include figure files
\usepackage{dcolumn}% Align table columns on decimal point
\usepackage{bm}% bold math
\begin{document}

%\preprint{APS/123-QED}

\title{\textbf{Beyond the Quasi-Steady Approximation: Modeling Unsteady Flows from Steady Solution Families.} 
}% 

\author{Takenori Shimozono}
\affiliation{Department of Civil Engineering, The University of Tokyo, Tokyo, Japan
}%

\date{\today}% It is always \today, today,
             
\begin{abstract}
Many flow systems have well-established steady-state relations, but applying them quasi-steadily under unsteady forcing neglects memory effects. This study develops a modeling framework extending steady-state relations to unsteady flows by closing dynamic departures from the instantaneous steady state. Integral conservation laws determine the explicit balance structure, while finite-order temporal closures approximate unresolved spatial dynamics responsible for the memory effect. This formulation preserves the prescribed nonlinear steady-state relations as equilibria under constant forcing. Canonical benchmarks in laminar and turbulent channel flows and shallow-water wave propagation demonstrate that the required temporal structure depends on spatial memory and forcing timescale. First-order relaxation captures viscous diffusion, whereas wave propagation and boundary reflections require higher-order models for oscillatory memory. Modal analysis further reveals that the required dynamic resolution depends on the predicted quantity, with wall shear being more sensitive to higher viscous modes than bulk flow rate. Nonlinear extensions with state-dependent steady relations and dynamic coefficients reproduce transient lag, relaxation, and oscillations under slow-to-moderate forcing, including finite-amplitude regimes. However, rapid finite-amplitude waves expose the limits of locally linearized closures in capturing nonlinear phase dynamics. This framework compactly extends quasi-steady approximations while preserving exact steady-state consistency.
\end{abstract}

\maketitle
\newpage

%\tableofcontents
\section{Introduction}
While the steady-state behavior of many fluid dynamical systems is well understood through analytical solutions or empirical formulations, extending such knowledge to unsteady conditions remains a modeling challenge. In engineering practice, a quasi-steady approximation---in which the instantaneous response is evaluated directly from a steady-state relation---is frequently employed under weakly unsteady conditions. However, as the forcing timescale approaches the characteristic response time of the system, phase lags and memory effects become increasingly important \cite{Lighthill1954,Womersley1955}, and the quasi-steady approximation deteriorates. Although data-driven surrogates and operator-learning techniques have recently emerged to construct input--output mappings under general time-dependent forcing \cite{Brunton2020,Lu2021}, accurate unsteady prediction for complex geometries or strongly nonlinear flows still largely relies on full numerical simulations that resolve the spatio-temporal dynamics governed by partial differential equations (PDEs), at substantial computational cost. 

Related approaches to representing unresolved spatial dynamics have been developed in various forms. Classical reductions such as depth averaging and cross-sectional integration eliminate selected spatial degrees of freedom by representing their effects in an averaged form. For systems with more complex spatial structure, projection-based model order reduction techniques, most notably Proper Orthogonal Decomposition coupled with Galerkin projection (POD--Galerkin), are widely used to approximate full-order dynamics \cite{Berkooz1993,Noack2003,Holmes2012,Rowley2017}. Regardless of the reduction strategy, eliminating spatial degrees of freedom generally requires a closure for the unresolved dynamics. In non-Markovian processes, these effects take the form of time-convolution memory integrals---a mechanism long recognized in classical analytical and semi-analytical formulations across fluid mechanics \cite{Basset1888,Hayami1951,Cummins1962,Zielke1968,Shimozono2020}, and formally generalized by the Mori--Zwanzig (MZ) framework \cite{Mori1965,Zwanzig1961,Chorin2002,ParishDuraisamy2017}. In many physical domains, such nonlocal memory representations have in turn been approximated by finite-order ordinary differential equation (ODE) models \cite{Oldroyd,Dooge1973,Trikha1975,LeishmanBeddoes1989,Peters1995,OlufsenNadim2004,Brunton2013}.

Since steady-state relations and integral conservation laws are already available for many engineering systems, an unsteady model need not be constructed entirely from scratch. Instead, this prior physical knowledge can determine part of the model structure, restricting the closure problem to the remaining unresolved contributions. Although reduced-order and data-driven approaches have been widely developed for unsteady flows, these models do not generally guarantee satisfaction of the underlying conservation laws unless appropriate constraints are explicitly enforced \cite{Carlberg2018}. Preserving an established steady-state relation is another important requirement in practical engineering applications. These considerations motivate building the model from the steady-state side, extending known steady-state relations through the quasi-steady regime and beyond. While conventional quasi-steady approximations and low-frequency algebraic corrections preserve the nonlinear steady response, they cannot capture internal relaxation, wave propagation, or spatial memory under faster forcing. The modeling challenge is thus to incorporate these dynamic effects while retaining both the prescribed steady-state relation and the balance structure imposed by integral conservation laws.

The objective of this work is to extend steady-state relations beyond the quasi-steady approximation while preserving their nonlinear structure. Rather than evaluating the bulk response of interest, such as flow rates or net forces, solely from the instantaneous steady solution, this approach uses that solution as a moving reference and constructs reduced-order dynamics for departures from it. The unsteady departures arising from unresolved spatial dynamics are captured by finite-order temporal closures, while integral conservation laws govern the separation of explicit and closure contributions. The resulting framework yields time-local ODE models that remain anchored to the nonlinear steady-state relations and capture transient effects beyond the quasi-steady regime. This study further examines how the applicability of the framework and the required temporal closure structure depend on the underlying physical mechanisms and the choice of predicted quantity.

Because the framework is anchored to a steady-state family, it is intended for systems that remain dynamically connected to a smooth family of steady states and whose unresolved spatial dynamics admit an adequate finite-order temporal representation. Its applicability may therefore become limited when discontinuous or naturally unstable behavior dominates. Within this scope, the present work evaluates the framework using two contrasting classes of PDE systems: plane Poiseuille flow as a parabolic viscous-diffusion system and finite-length Saint--Venant flow as a hyperbolic wave-propagation system. These canonical benchmarks demonstrate how distinct spatial mechanisms give rise to different temporal structures after spatial reduction. The framework is then extended in both cases to nonlinear systems with state-dependent steady solutions and local dynamic properties to assess its performance for finite-amplitude transients across a range of forcing timescales.

\section{Theoretical Framework}
We consider the general problem of predicting the bulk response $\tilde{Y}(\tilde{t})$ subjected to dynamic forcing $\tilde{\lambda}(\tilde{t})$, where a tilde denotes a dimensional quantity. The present framework considers dissipative fluid systems that naturally relax toward a well-defined family of steady states $\tilde{Y}_s(\tilde{\lambda})$ when forcing unsteadiness vanishes. We assume a smooth steady-state family over the operating range, excluding bifurcation and self-excited instability. Under dynamic conditions, internal relaxation dynamics introduce spatial memory effects, causing the actual response $\tilde{Y}(\tilde{t})$ to depart from the instantaneous steady state $\tilde{Y}_s(\tilde{\lambda}(\tilde{t}))$. We quantify this dynamic departure by
\begin{equation}
\Delta\widetilde{Y}(\widetilde{t}) = \widetilde{Y}(\widetilde{t}) - \widetilde{Y}_s(\widetilde{\lambda}(\widetilde{t})).
\label{eq:dim_departure}
\end{equation}

Assuming this dynamic departure remains bounded and sufficiently smooth over the regime of interest, we construct a low-order dynamical model anchored by the integral conservation law,
\begin{equation}
\dot{\widetilde{I}} = \widetilde{R},
\label{eq:dim_reynolds}
\end{equation}
where $\widetilde{I}$ represents the integral storage of the conserved quantity and $\widetilde{R}$ denotes its net rate of supply across the control volume. Thus, the conservation law is used as the physical constraint coupling the dynamic response $\Delta\widetilde{Y}(\widetilde{t})$ to the external forcing $\widetilde{\lambda}(\widetilde{t})$.

To formulate a nondimensional theory, we introduce characteristic scales associated with the system: a characteristic response and forcing timescale ($T_r$, $T_f$), a reference scale $\lambda_r$ for the forcing parameter, an integral storage scale $I_r$, and a representative response scale $Y_r$, which may be chosen from the steady-state family. Defining the slow-rate parameter as
\begin{equation}
\epsilon = \frac{T_r}{T_f},
\end{equation}
we nondimensionalize time as $t = \widetilde{t} / T_f$ and scale all variables according to
\begin{equation}
\lambda = \frac{\widetilde{\lambda}}{\lambda_r}, \qquad Y = \frac{\widetilde{Y}}{Y_r}, \qquad I = \frac{\widetilde{I}}{I_r}, \qquad R = \frac{\widetilde{R} \, T_r}{I_r}.
\label{eq:nondim_scales}
\end{equation}
Here, the choice of $T_r$ rather than $T_f$ for scaling $\widetilde R$ naturally aligns with the internal relaxation mechanics of the system.

Defining the dimensionless steady-state family as $Y_s(\lambda(t)) = \widetilde{Y}_s(\widetilde{\lambda}(\widetilde{t})) / Y_r$, the dimensionless departure corresponding to Eq.~\eqref{eq:dim_departure} becomes $\Delta Y(t) = Y(t) - Y_s(\lambda(t))$, where $\Delta Y = 0$ indicates that the selected bulk response coincides with its instantaneous steady value, but does not necessarily imply that the full spatial state is steady. 

Under these scalings, the integral conservation law in Eq.~\eqref{eq:dim_reynolds} transforms into the singularly perturbed form:
\begin{equation}
\epsilon \dot{I} = R,
\label{eq:reynolds_nondim}
\end{equation}
where
\begin{equation}
I = \int_{\Omega(t)} \psi \, \mathrm{d}V, \qquad R = -\int_{\partial\Omega(t)} \psi (\mathbf{u} - \mathbf{u}_{\mathrm{CS}}) \cdot \mathbf{n} \, \mathrm{d}A + R_b + R_v.
\label{eq:reynolds_integrals}
\end{equation}
Here, $\Omega(t)$ denotes the time-dependent dimensionless control volume bounded by surface $\partial\Omega(t)$ with outward unit normal $\mathbf{n}$. The quantity $\psi$ is the normalized conserved density, $\mathbf{u}$ and $\mathbf{u}_{\mathrm{CS}}$ are the normalized fluid and control-surface velocity vectors, while $R_b$ and $R_v$ denote dimensionless boundary fluxes (including surface forces) and volumetric body forces, respectively.

In the steady limit ($\epsilon \to 0$), the steady-state solution satisfies the exact integral balance, so that the net supply rate vanishes along the family of steady states, $R_s(\lambda)\equiv 0$. Under weakly unsteady forcing ($\epsilon \ll 1$), the departure from this balance is small, with $R=\epsilon \dot{I}=\mathcal{O}(\epsilon)$.  This scaling motivates organizing the temporal closure in powers of the scaled time-differentiation operator $\epsilon\,\mathrm d/\mathrm dt$, without imposing a prescribed amplitude on the dynamic departure.

To separate the instantaneous steady baseline from the dynamic departure, we first expand the integral storage as
\begin{equation}
I(t)=I_s(\lambda(t))+\Delta I(t),
\label{eq:I_decomposition}
\end{equation}
where $I_s(\lambda)$ denotes the storage along the steady-state family. Since the net supply rate vanishes on this family, $R_s(\lambda)=0$, the exact integral conservation law takes the departure form
\begin{equation}
\epsilon I_s'(\lambda)\dot\lambda+\epsilon \dot{\Delta I}=R.
\label{eq:integral_departure_balance}
\end{equation}
The first term explicitly accounts for the forcing-induced motion of the steady reference state, while the remaining terms ($\epsilon \dot{\Delta I}$ and $R$) describe departures from its instantaneous steady integral balance. Such departures may reflect the deformation of unresolved spatial profiles within the control volume.

We next separate the departure terms that can be evaluated explicitly using $Y$ and $\lambda$ from those that still depend on unresolved spatial information. Specifically, with $\Delta Y=Y-Y_s(\lambda)$, we write the exact rearrangement
\begin{equation}
\begin{aligned}
\Delta I &= \Delta I_{\mathrm{res}}(\Delta Y,\lambda)+\Delta I_{\mathrm{unres}},\\
R &= R_{\mathrm{res}}(\Delta Y,\lambda)+R_{\mathrm{unres}},
\end{aligned}
\label{eq:resolved_unresolved_decomposition}
\end{equation}
where $\Delta I_{\mathrm{res}}$ and $R_{\mathrm{res}}$ contain the explicitly resolved departure contributions, and $\Delta I_{\mathrm{unres}}$ and $R_{\mathrm{unres}}$ contain the remaining spatially unresolved contributions that require closure. The resolved departure terms are chosen such that $\Delta I_{\mathrm{res}}(0,\lambda)=0$ and $R_{\mathrm{res}}(0,\lambda)=0$. Their specific expressions depend on exact conservation identities and the specific bulk output of interest. This rearrangement is exact and introduces no approximations at this stage.

Substituting Eq.~\eqref{eq:resolved_unresolved_decomposition} into Eq.~\eqref{eq:integral_departure_balance} gives
\begin{equation}
\epsilon I_s'(\lambda)\dot{\lambda}+\epsilon \dot{\Delta I}_{\mathrm{res}}(\Delta Y,\lambda)-R_{\mathrm{res}}(\Delta Y,\lambda)=\mathcal U,
\label{eq:combined_unresolved_balance}
\end{equation}
where
\begin{equation}
\mathcal U\equiv R_{\mathrm{unres}}-\epsilon \dot{\Delta I}_{\mathrm{unres}}
\label{eq:combined_unresolved_term}
\end{equation}
is the combined unresolved contribution to the departure balance. All explicitly evaluable terms, including nonlinear terms when available from exact integral identities, remain on the resolved left-hand side. Only the remaining unresolved contribution $\mathcal U$ requires a closure model. The construction is consistent with the steady-state family: for a spatially steady state under constant forcing ($\dot{\lambda}=0$), $\Delta Y=0$ and $\mathcal U=0$.

We approximate this combined departure contribution by a finite-order temporal closure around the moving steady-state baseline:
\begin{equation}
\mathcal U\simeq\sum_{k=0}^{M}\epsilon^k C_k(\lambda)\Delta Y^{(k)}+\sum_{k=1}^{N}\epsilon^k D_k(\lambda)\lambda^{(k)}.
\label{eq:U_general_nondim}
\end{equation}
Here, $\Delta Y^{(k)}=\mathrm d^k\Delta Y/\mathrm dt^k$ and $\lambda^{(k)}=\mathrm d^k\lambda/\mathrm dt^k$. The departure series allows a zeroth-order term proportional to $\Delta Y$, since this term vanishes on the steady-state family. In contrast, the forcing series begins at $k=1$, because a term depending only on $\lambda$ would generally remain nonzero under constant forcing and violate the requirement that $\mathcal U=0$ for a spatially steady state. The powers of $\epsilon$ organize the closure in terms of the scaled time-differentiation operator $\epsilon\,\mathrm d/\mathrm dt$, without imposing a prescribed amplitude scaling on $\Delta Y$. The closure order therefore specifies the highest retained temporal derivative, rather than the truncation order of a regular perturbation expansion of the transient solution. The retained derivatives constitute a finite-order dynamic approximation, including terms that may be formally higher order for a particular slow-response solution. The unresolved contribution is approximated locally by a linear combination of departure and forcing derivatives, with coefficients that may vary nonlinearly along the steady-state family. This local linear representation is a modeling approximation rather than a consequence of the $\epsilon$ ordering. Nonlinear terms explicitly determined by the conservation law are retained in the resolved balance, while nonlinear interactions within the unresolved contribution are not modeled explicitly.

Substituting Eq.~\eqref{eq:U_general_nondim} into Eq.~\eqref{eq:combined_unresolved_balance} yields the governing ODE
\begin{equation}
\epsilon I_s'(\lambda)\dot\lambda+\epsilon\dot{\Delta I}_{\mathrm{res}}(\Delta Y,\lambda)-R_{\mathrm{res}}(\Delta Y,\lambda)\simeq\sum_{k=0}^{M}\epsilon^k C_k(\lambda)\Delta Y^{(k)}+\sum_{k=1}^{N}\epsilon^k D_k(\lambda)\lambda^{(k)}.
\label{eq:general_ode_nondim}
\end{equation}
This formulation retains the steady-state family and the explicitly determined departure balance, while approximating only the unresolved dynamic contribution. If the resolved storage departure is algebraic in $(\Delta Y,\lambda)$ and the resulting initial-value problem is well-posed, the dynamic order is at most $p=\max(1,M)$, with the actual order determined by the highest retained derivative of $\Delta Y$. While the conventional quasi-steady approximation neglects dynamic departures and evaluates the response directly as $Y_s(\lambda(t))$, the present model evolves $\Delta Y$ about this moving reference. This enables the model to capture transient relaxation and higher-order dynamic effects, given an appropriate dynamic order, while retaining the same steady-state relation.

The proposed framework does not prescribe a particular method for determining the temporal closure coefficients. These coefficients may be derived analytically from the governing physics, inferred from steady-state or perturbation calculations, or calibrated against a limited set of unsteady simulations, depending on the system and the information available. In practice, the truncation orders $M$ and $N$ depend on both the smallness of $\epsilon$ and the level of dynamic memory to be captured. The closure coefficients can be identified from the low-frequency response associated with $\epsilon \ll1$, and the resulting finite-order dynamic model is integrated directly as an ODE. The benchmarks below also examine larger values of $\epsilon$ to assess the closure performance beyond the slow-response regime.

In the following sections, we evaluate the framework using two canonical benchmark problems representing parabolic and hyperbolic dynamics, respectively. These benchmarks are used to examine how the required temporal closure structure depends on the underlying spatial dynamics and forcing timescale.
%==========================================================================================
\section{Canonical Validation I: Viscous-Diffusion Dynamics}
%==========================================================================================
The first canonical validation of the present framework evaluates viscous-diffusion dynamics in channel flow. The analysis begins with unsteady plane Poiseuille flow, for which the governing equations permit an exact characterization of the steady-state family and dynamic response. The reduction structure is subsequently extended to unsteady turbulent channel flow, where state-dependent eddy transport introduces nonlinear momentum diffusion. This tests whether the framework remains robust when the steady-state family and dynamic closure coefficients vary nonlinearly with the forcing state.
%==========================================================================================
\subsection{Governing system, steady-state family, and momentum balance}
Let us consider a symmetric, incompressible viscous flow between parallel plates located at nondimensional wall-normal coordinates $y = \pm1$, scaled by the channel half-width $a$. Time is nondimensionalized by the characteristic forcing timescale $T_f$, introducing the dynamic timescale ratio $\epsilon = T_r/T_f$, where $T_r = a^2/\nu$ is the intrinsic viscous diffusion timescale. The dimensional velocity and volumetric flow rate per unit span are scaled by $q_r/a$ and $q_r$, respectively, where $q_r$ is a representative dimensional flow-rate scale. Under these scalings, the nondimensional pressure gradient $F(t)$ serves as the prescribed forcing, while the bulk response of interest is the nondimensional volumetric flow rate per unit span $q(t)$, defined as

\begin{equation}
q(t) = \int_{-1}^{1} u(y,t) \, \mathrm{d}y,
\end{equation}
where $u(y,t)$ is the nondimensional streamwise velocity field. Assuming fully developed laminar flow, $u(y,t)$ satisfies the dimensionless singularly perturbed diffusion equation,

\begin{equation}
\epsilon \frac{\partial u}{\partial t} = F(t) + \frac{\partial^2 u}{\partial y^2}, 
\label{eq:nondim_poiseuille}
\end{equation}
subject to no-slip boundary conditions $u(\pm 1, t) = 0$ and the quiescent initial condition $u(y,0)=0$.

Predicting $q(t)$ directly from $F(t)$ conventionally requires solving the governing partial differential equation. For a constant forcing $F$, the steady velocity profile and corresponding flow rate are

\begin{equation}
u_s(y;F) = \frac{F}{2}(1-y^2), \qquad q_s(F) = \frac{2}{3}F,
\label{eq:steady_poiseuille}
\end{equation}
which define the steady-state family. The corresponding steady differential response with respect to forcing is constant and given by $q_s'(F) \equiv \mathrm{d}q_s/\mathrm{d}F = 2/3$.

Integrating Eq.~\eqref{eq:nondim_poiseuille} across the non-dimensional channel width yields the dimensionless momentum balance:
\begin{equation}
\epsilon \dot{q} = 2F - 2\tau_w, \quad \text{where} \quad \tau_w(t) = -\left. \frac{\partial u}{\partial y} \right|_{y = 1}.
\label{eq:momentum}
\end{equation}
In the notation of Section~II, $Y=I=q$, $\lambda=F$ and $R= 2F - 2\tau_w$. Thus, the integral storage is directly determined by the resolved output, whereas the wall shear stress $\tau_w$ remains unresolved because it depends on the near-wall velocity profile.

Next, we decompose $q(t)$ and $\tau_w(t)$ into their values along the family of steady states and their dynamic departures:
\begin{equation}
q(t) = q_s(F(t)) + \Delta q(t), \qquad \tau_w(t) = \tau_{w,s}(F(t)) + \Delta \tau_w(t),
\label{eq:departure}
\end{equation}
where $\tau_{w,s}(F) = F$ represents the wall shear stress required for steady momentum equilibrium ($R = 0$). The nondimensional dynamic perturbations $\Delta q(t)$ and $\Delta \tau_w(t)$ quantify departures from the instantaneous steady state and vanish identically under steady-state conditions.

Substituting Eq.~\eqref{eq:departure} into Eq.~\eqref{eq:momentum} and using the steady equilibrium relation $\tau_{w,s}(F)=F$, we obtain
\begin{equation}
\epsilon \dot{q} =\epsilon q_s'(F)\dot{F} + \epsilon \dot{\Delta q} = -2\Delta\tau_w.
\label{eq:momentum_dimensionless}
\end{equation}
This equation governs the dynamic coupling between the flow-rate departure $\Delta q(t)$, the wall-shear stress departure $\Delta \tau_w(t)$, and the forcing rate $\dot{F}(t)$. The only unresolved contribution is the dynamic flux departure $-2\Delta\tau_w$.
%=========================================
\subsection{Low-order dynamic closure and reduced model}
Following the combined temporal-closure representation in Eq.~\eqref{eq:U_general_nondim}, the unresolved contribution to the integral balance is $\mathcal U=-2\Delta\tau_w$. We adopt a temporal closure retaining forcing derivatives through second order:
\begin{equation}
\mathcal U \simeq C_0 \Delta q + \epsilon C_1 \dot{\Delta q} + \epsilon^2 C_2 \ddot{\Delta q} + \epsilon D_1 \dot{F} + \epsilon^2 D_2 \ddot{F}.
\label{eq:poiseuille_R_general}
\end{equation}
The unresolved dynamics are contained entirely in the closure coefficients $(C_0, C_1, C_2, D_1, D_2)$.
Substituting Eq.~\eqref{eq:poiseuille_R_general} into the exact departure balance, Eq.~\eqref{eq:momentum_dimensionless}, gives
\begin{equation}
\epsilon(1 -  C_1) \dot{\Delta q} - \epsilon^2 C_2 \ddot{\Delta q} \simeq C_0 \Delta q + \epsilon \left[ D_1 - q_s'(F) \right] \dot{F} + \epsilon^2 D_2 \ddot{F}.
\label{eq:poiseuille_R1_balance}
\end{equation}
Provided $1 - C_1 \neq 0$, the coefficient $C_1$ does not introduce an additional dynamic degree of freedom, but only rescales the effective relaxation timescale and remaining coefficients through the factor $1 - C_1$. We therefore choose the normalization $C_1 = 0$ and rewrite Eq.~\eqref{eq:poiseuille_R1_balance} as
\begin{equation}
- \epsilon^2 C_2 \ddot{\Delta q} + \epsilon \dot{\Delta q} \simeq C_0 \Delta q + \epsilon \left[ D_1 - q_s'(F) \right] \dot{F} + \epsilon^2 D_2 \ddot{F}.
\label{eq:poiseuille_departure_ode}
\end{equation}
Expressing this in terms of the total flow rate predicted by the reduced model, $q_{\mathrm{ode}}(t) \equiv q_s(F(t)) + \Delta q(t)$, yields the equivalent form:
\begin{equation}
- \epsilon^2 C_2 \ddot{q}_{\mathrm{ode}} + \epsilon \dot{q}_{\mathrm{ode}} = C_0 \left[ q_{\mathrm{ode}} - q_s(F) \right] + \epsilon D_1 \dot{F} + \epsilon^2 \left[ D_2 - C_2 q_s'(F) \right] \ddot{F}.
\label{eq:poiseuille_total_ode}
\end{equation}
The corresponding wall shear stress follows directly from the exact momentum balance as $\tau_{w,\mathrm{ode}} = F - \epsilon \dot{q}_{\mathrm{ode}}/2$. For plane Poiseuille flow, the governing momentum equation is linear and the steady-state relation is simply $q_s(F) = 2F/3$ (yielding $q_s'(F) = 2/3$). The dynamic response around this baseline is independent of the forcing level $F$. The closure coefficients $(C_0, C_2, D_1, D_2)$ thus reduce to state-independent constants.
An explicit algebraic approximation follows by asymptotically expanding Eq.~\eqref{eq:poiseuille_total_ode} in terms of $\epsilon$. Retaining terms through $\mathcal{O}(\epsilon^2)$ gives
\begin{equation}
q_{\mathrm{eqs}}(t) = q_s(F) + \epsilon \left[ \frac{2- 3 D_1}{3C_0} \right] \dot{F} + \epsilon^2 \left[ \frac{2 -3 D_1}{3 C_0^2} - \frac{D_2}{C_0} \right] \ddot{F} + \mathcal{O}(\epsilon^3).
\label{eq:approx}
\end{equation}
This algebraic approximation $q_{\mathrm{eqs}}(t)$ is referred to hereafter as the extended quasi-steady (EQS) solution. Retaining only the leading-order term ($\mathcal{O}(1)$) recovers the classical quasi-steady (QS) approximation, $q_s(F(t))$. This closed-form approximation provides insight into perturbations due to unsteady effects. For general time-varying forcing, the $\dot F$ term represents a rate-dependent response (or temporal lag), whereas the $\ddot F$ term accounts for acceleration effects on local extrema (such as peak enhancement or reduction).
%==========================================================================================
\subsection{Linear response and coefficient identification}
We next consider harmonic forcing of the linear system,
\begin{equation}
F(t) = \mathrm{Re}\{\widehat{F} e^{i\omega t}\},
\label{eq:harmonic_forcing}
\end{equation}
where $\widehat{F}$ and $\omega = \tilde{\omega} T_f$ denote the complex amplitude and dimensionless forcing frequency, respectively. Introducing the intrinsic-time-scaled complex frequency $\sigma = i \epsilon \omega = i \tilde{\omega}T_r$ corresponding to the dimensionless time $t$, the exact transfer functions from $F$ to $q$, normalized by their respective steady-state sensitivities $q_s'(F)$ and $\tau_{w,s}'(F)$, are expressed as
\begin{equation}
\begin{aligned}
G_q(\sigma) &\equiv \frac{\widehat{q}(\sigma)}{q_s'(F) \widehat{F}(\sigma)} = \frac{3}{\sigma} \left( 1 - \frac{\tanh\sqrt{\sigma}}{\sqrt{\sigma}} \right), \\
G_\tau(\sigma) &\equiv \frac{\widehat{\tau}_w(\sigma)}{\tau_{w,s}'(F) \widehat{F}(\sigma)} = 1 - \frac{\sigma}{3} G_q(\sigma) = \frac{\tanh\sqrt{\sigma}}{\sqrt{\sigma}}.
\label{eq:transfer_funcs}
\end{aligned}
\end{equation}
Both transfer functions satisfy the quasi-steady limit $G_q(0) = G_\tau(0) = 1$. Their transcendental form reflects the infinite spectrum of viscous-diffusion modes and therefore cannot be reproduced over all frequencies by a finite-order ODE.

Similarly, the corresponding normalized flow-rate and wall-stress transfer functions for the $p$th order ODE model are obtained from Eq.~\eqref{eq:poiseuille_total_ode}:
\begin{equation}
G^{\mathrm{ode}}_{q,p} (\sigma) = \frac{{ - C_{0,p}  + \frac{3}{2}D_{1,p} \sigma+ \left( {\frac{3}{2}D_{2,p}  - C_{2,p} } \right)\sigma^2 }}{{ - C_{0,p}  + \sigma - C_{2,p} \sigma^2 }},\qquad G^{\mathrm{ode}}_{\tau ,p} (\sigma) = 1 - \frac{\sigma}{3}G^{\mathrm{ode}}_{q,p} (\sigma).
\label{eq:G_ode}
\end{equation}
Here, $G^{\mathrm{ode}}_{q,p}(\sigma)$ and $G^{\mathrm{ode}}_{\tau,p}(\sigma)$ represent the rational transfer functions of dynamic order $p$. For the first-order closure ($p=1$), we set $C_{2,1} = D_{2,1} = 0$. Because the ODE model represents a transcendental operator with infinite poles using a finite rational approximant, the optimal closure coefficients depend explicitly on the order $p$.

The EQS transfer functions  corresponding to Eq.~\eqref{eq:approx} are
\begin{equation}
G^{\mathrm{eqs}}_{q,p} (\sigma) = 1 + \frac{{2 - 3D_{1,p} }}{{2C_{0,p} }}\sigma + \left[ {\frac{{2 - 3D_{1,p} }}{{2C_{0,p}^2 }} - \frac{{3D_{2,p} }}{{2C_{0,p} }}} \right]\sigma^2 ,\qquad G^{\mathrm{eqs}}_{\tau ,p} (\sigma) = 1 - \frac{\sigma}{3}G^{\mathrm{eqs}}_{q,p} (\sigma).
\label{eq:G_eqs}
\end{equation}
The EQS approximation is obtained by expanding the ODE transfer function about $\sigma=0$ and retaining terms through second order. It therefore contains no relaxation poles and is restricted to the low-frequency regime.

The exact flow-rate transfer function has the low-frequency expansion
\begin{equation}
G_q(\sigma) = 1 - \frac{2}{5}\sigma + \frac{17}{105}\sigma^2 - \frac{62}{945}\sigma^3 + \frac{1382}{51975}\sigma^4 + \mathcal{O}(\sigma^5).
\label{eq:Gq_exact_taylor}
\end{equation}
Expanding Eq.~\eqref{eq:G_ode} about $\sigma = 0$ and matching Eq.~\eqref{eq:Gq_exact_taylor} through $\mathcal{O}(\sigma^{2p})$ uniquely determines the dimensionless closure coefficients as
\begin{equation}
\begin{aligned}
p=1: \qquad \left( {C_{0,1} ,C_{2,1} ,D_{1,1} ,D_{2,1} } \right) &= \left( { - \frac{{42}}{{17}},0,\frac{2}{{255}},0} \right),\\ 
p=2: \qquad \left( {C_{0,2} ,C_{2,2} ,D_{1,2} ,D_{2,2} } \right) &= \left( { - \frac{{165}}{{74}}, - \frac{{13}}{{333}},\frac{8}{{111}}, - \frac{{101}}{{3885}}} \right)
\end{aligned}
\label{eq:q_ode_coefficients}
\end{equation}
Because the reconstructed wall shear satisfies $G^{\mathrm{ode}}_{\tau,p}(\sigma) = 1 - \sigma G^{\mathrm{ode}}_{q,p}(\sigma)/3$, matching $G^{\mathrm{ode}}_{q,p}(\sigma)$ through $\mathcal{O}(\sigma^{2p})$ automatically matches $G^{\mathrm{ode}}_{\tau,p}(\sigma)$ through $\mathcal{O}(\sigma^{2p+1} )$; no independent wall-shear coefficients are required.
\begin{figure}[t]
\centering
\includegraphics[width=1\linewidth]{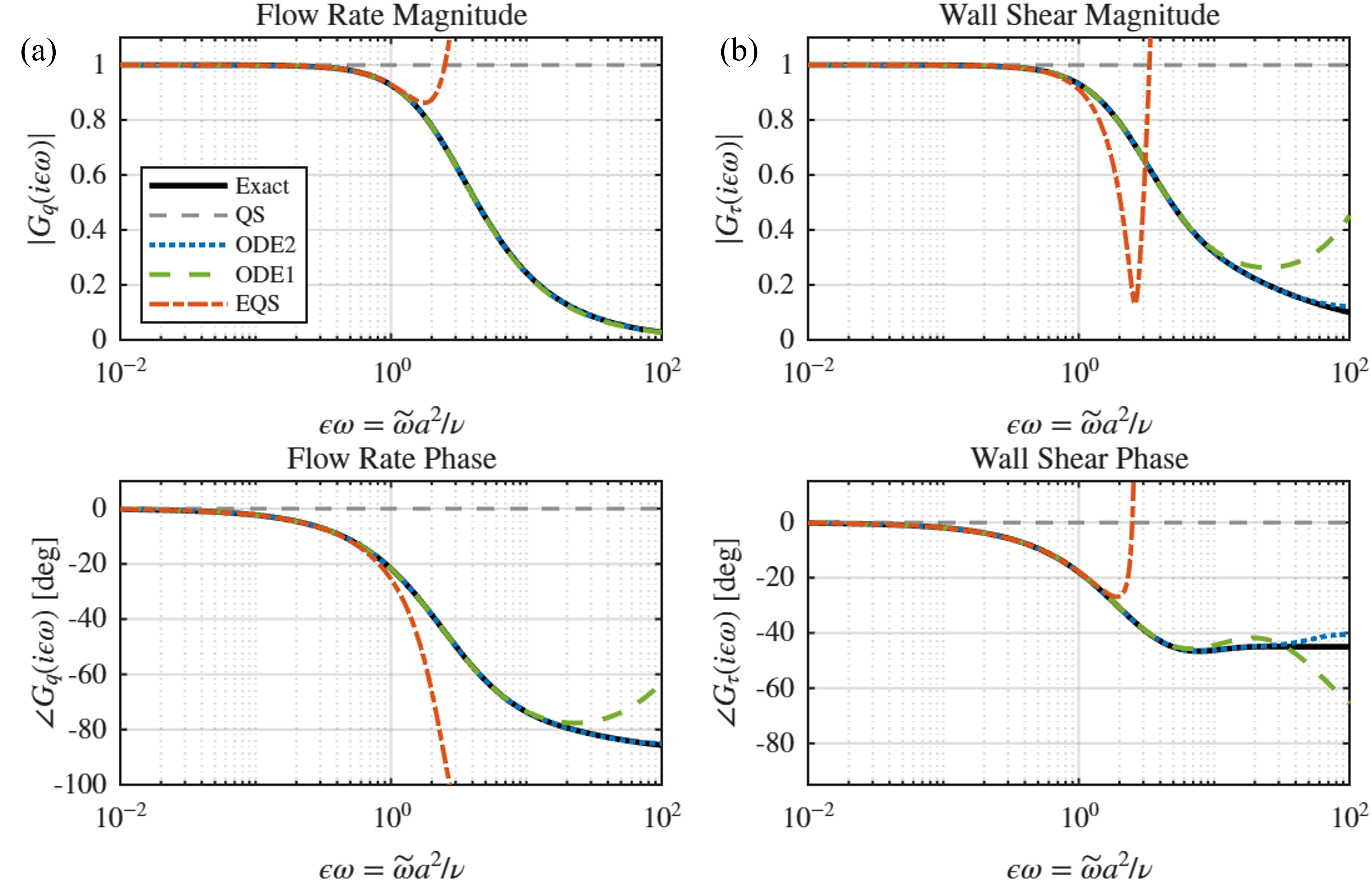}
\caption{Comparison of transfer-function magnitude $|G(i\epsilon \omega)|$ (top) and phase response $\angle G(i\epsilon \omega)$ (bottom) under harmonic forcing across a range of dimensionless frequencies $\omega = \tilde{\omega} T_f$ (intrinsic frequency $\epsilon \omega = \tilde{\omega} a^2/\nu$): (a) volumetric flow rate $q$ and (b) reconstructed wall shear stress $\tau_w$. Results are compared among the exact solution, quasi-steady (QS) approximation, extended quasi-steady (EQS) approximation, the ODE models with $p=1$(ODE1) and $p=2$(ODE2).}
\label{fig:1}
\end{figure}

Figure~\ref{fig:1} compares the magnitude and phase responses of the reduced models with the exact solution. All models recover the quasi-steady limit as $\omega \to 0$. The EQS approximation improves upon the classical QS response at low frequencies but deteriorates as $\epsilon \omega$ approaches $\mathcal{O}(1)$, where the local expansion ceases to be accurate. On the other hand, both ODE models retain the relaxation dynamics and reproduce the exact flow-rate magnitude and phase over a substantially broader frequency range. The $p=2$ ODE model exhibits better agreement with the exact solution than the $p=1$ model at higher frequencies, demonstrating the benefit of incorporating higher-order temporal derivatives in capturing unsteady departures.

The high-frequency behavior further illustrates the fundamental distinction among different models. As $\epsilon \omega \to \infty$, the exact transcendental solution exhibits the characteristic asymptotic roll-off $G_q(\sigma) \sim 3(\sigma)^{-1}$ and $G_\tau(\sigma) \sim (\sigma)^{-1/2}$, reflecting viscous diffusion confined to a thin Stokes boundary layer. In contrast, the EQS approximation lacks a dynamic relaxation pole and diverges polynomial-wise as $\mathcal{O}(\sigma^2)$ in the high-frequency limit. The ODE models, on the other hand, maintain strictly bounded flow-rate responses: as $\epsilon \omega \to \infty$, the transfer functions approach finite asymptotic limits $(G^{\mathrm{ode}}_{q,1}, G^{\mathrm{ode}}_{q,2}) \to (1/85, 1/910)$ for $p=1$ and $p=2$, respectively. The reconstructed wall shear stress nevertheless remains sensitive to high-frequency forcing through the exact momentum balance, $G^{\mathrm{ode}}_{\tau,p}(\sigma) = 1 - \sigma G^{\mathrm{ode}}_{q,p}(\sigma)/3$, which directly transmits high-order forcing derivatives to the near-wall velocity gradient.
%======================================================
\subsection{Transient validation}
We next evaluate the time-domain performance of the ODE models using the matched coefficients in Eq.~\eqref{eq:q_ode_coefficients}. Only the primary flow-rate equation is integrated, with the wall shear stress reconstructed directly from the exact momentum balance.
The exact transient responses can be written in convolution form as
\begin{equation}
q(t) = \int_0^t K_q(t-\xi) \dot{F}(\xi) \, \mathrm{d}\xi, \qquad \tau_w(t) = \int_0^t K_\tau(t-\xi) \dot{F}(\xi) \, \mathrm{d}\xi,
\label{eq:conv_exact_system}
\end{equation}
where
\begin{equation}
K_q(t) = \frac{2}{3} \left( 1 - \sum_{n=0}^{\infty} a_{q,n} e^{-\gamma_n t / \epsilon} \right), \qquad K_\tau(t) = 1 - \sum_{n=0}^{\infty} a_{\tau,n} e^{-\gamma_n t / \epsilon},
\label{eq:exact_kernels_poiseuille}
\end{equation}
with
\begin{equation}
a_{q,n} = \frac{96}{\pi^4(2n+1)^4}, \qquad a_{\tau,n} = \frac{8}{\pi^2(2n+1)^2}, \qquad \gamma_n = \frac{(2n+1)^2\pi^2}{4}.
\label{eq:modal_parameters}
\end{equation}
Since $\sum_{n=0}^{\infty} a_{q,n} = \sum_{n=0}^{\infty} a_{\tau,n} = 1$, the kernels satisfy $K_q(0) = K_\tau(0) = 0$ and approach $K_q(\infty) = 2/3$ and $K_\tau(\infty) = 1$.
Equations~\eqref{eq:conv_exact_system}--\eqref{eq:modal_parameters} represent an infinite hierarchy of viscous-diffusion modes decaying at rates $\gamma_n/\epsilon$. Traditional formulations of transient boundary shear similarly involve time-history representations \citep{LiuOrfila2004,Liu2006}, whereas the ODE models approximate the distributed memory using a finite number of local dynamic states.

For quiescent initial conditions, the first-order ODE model admits the convolution kernels
\begin{equation}
\begin{aligned}
K^{\mathrm{ode}}_{q,1}(t) &= \frac{2}{3} \left[ 1 - \left(1 - \frac{3}{2}D_{1,1}\right) e^{C_{0,1} t/\epsilon} \right], \\
K^{\mathrm{ode}}_{\tau,1}(t) &= 1 + \frac{C_{0,1}}{3} \left(1 - \frac{3}{2}D_{1,1}\right) e^{C_{0,1} t/\epsilon} - \frac{\epsilon D_{1,1}}{2} \delta(t),
\end{aligned}
\label{eq:ode1_kernels_poiseuille}
\end{equation}
where $\delta(t)$ is the Dirac delta function. The identified relaxation parameter $-C_{0,1} = 42/17 \approx 2.471$ closely matches the fundamental diffusion rate of the exact kernel $\gamma_0 = \pi^2/4 \approx 2.467$, indicating that the dominant reduced pole at $\sigma = C_{0,1}$ directly captures the slowest viscous-diffusion mode. Furthermore, the relaxation amplitudes for both variables, $1 - 3D_{1,1}/2 \approx 0.988$ and $-C_{0,1}(1 - \frac{3}{2}D_{1,1})/3 \approx 0.814$, align well with the fundamental modal weights of the exact system ($a_{q,0} = 96/\pi^4 \approx 0.986$ and $a_{\tau,0} = 8/\pi^2 \approx 0.811$). This confirms that the single reduced pole captures the dominant contribution of the fundamental viscous-diffusion mode in both flow rate and wall shear stress.

For the second-order closure ($p=2$), the explicit convolution kernels are given by
\begin{equation}
\begin{aligned}
K^{\mathrm{ode}}_{q,2}(t) &= \frac{2}{3} \left[ 1 - a_1 e^{-\kappa_1 t/\epsilon} - a_2 e^{-\kappa_2 t/\epsilon} \right], \\
K^{\mathrm{ode}}_{\tau,2}(t) &= 1 - \frac{1}{3} \left[ \kappa_1 a_1 e^{-\kappa_1 t/\epsilon} + \kappa_2 a_2 e^{-\kappa_2 t/\epsilon} \right] - \frac{\epsilon}{3} (1 - a_1 - a_2) \delta(t),
\end{aligned}
\label{eq:ode2_kernels_poiseuille}
\end{equation}
where the effective relaxation rates $\kappa_{1,2}$ depend exclusively on the state-closure coefficients $C_{0,2}$ and $C_{2,2}$ via
\begin{equation}
\kappa_{1,2} = \frac{-1 \pm \sqrt{1 - 4 C_{0,2} C_{2,2}}}{2 C_{2,2}} \qquad (\kappa_1 \approx 2.467, \,\, \kappa_2 \approx 23.15),
\label{eq:kappa_analytical}
\end{equation}
which closely track the first two intrinsic viscous-diffusion eigenvalues of the exact system ($\gamma_0 = \pi^2/4 \approx 2.467$ and $\gamma_1 = 9\pi^2/4 \approx 22.21$). The corresponding modal amplitude weights $a_1$ and $a_2$ combine both forcing ($D$) and state ($C$) closures according to
\begin{equation}
a_1 = \frac{\left(\frac{3}{2}D_{1,2} - 1\right) - \frac{3}{2}D_{2,2} \kappa_1}{C_{2,2}(\kappa_2 - \kappa_1)}, \qquad a_2 = \frac{-\left(\frac{3}{2}D_{1,2} - 1\right) + \frac{3}{2}D_{2,2} \kappa_2}{C_{2,2}(\kappa_2 - \kappa_1)},
\label{eq:modal_weights_analytical}
\end{equation}
yielding $a_1 \approx 0.9855$ and $a_2 \approx 0.0134$ with $a_1 + a_2 = 909/910 \approx 0.9989$, which indicates that nearly all of the flow-rate step-response amplitude is carried by the two relaxation modes.

The ODE models preserve the exact steady-state gains ($K_q(\infty) = 2/3$ and $K_\tau(\infty) = 1$) while replacing the infinite viscous-diffusion spectrum with a low-dimensional set of discrete relaxation modes decaying at rates $-C_{0,1}/\epsilon$ (for $p=1$) or $\kappa_{1,2}/\epsilon$ (for $p=2$). In both models, the nonzero high-frequency limits of the flow-rate transfer functions, $3D_{1,1}/2=1/85$ for $p=1$ and $1-a_1-a_2=1/910$ for $p=2$, give rise to singular impulse terms $\delta(t)$ in the reconstructed wall-shear kernels through the exact momentum balance. 

The non-zero initial flow-rate offsets, $K^{\mathrm{ode}}_{q,1}(0) = D_{1,1} \approx 0.00784$ and $K^{\mathrm{ode}}_{q,2}(0) = 1/1365 \approx 0.000733$, as well as nonzero initial values of the regular parts of the wall-shear kernels, $K^{\mathrm{ode}}_{\tau,1}(0^+) = 1 + C_{0,1}(1 - 3D_{1,1}/2)/3 \approx 0.186$ and $K^{\mathrm{ode}}_{\tau,2}(0^+) = 1 - (\kappa_1 a_1 + \kappa_2 a_2)/3 \approx 0.0863$ (accompanied by the singular distributions $-\epsilon \delta(t)/255$ and $-\epsilon \delta(t)/2730$, respectively), represent high-frequency artifacts. While the exact viscous-diffusion kernels satisfy $K_q(0) = K_\tau(0) = 0$ strictly from quiescent conditions, advancing the closure order from $p=1$ to $p=2$ reduces these short-time truncation artifacts, improving agreement with the exact response.

The transient response is evaluated using the exponential forcing,
\begin{equation}
F(t) = 1 - e^{-t},
\label{eq:exp_ramp}
\end{equation}
where the characteristic forcing timescale $T_f$ is identified with the physical rise time of the ramp. Under this choice, varying the timescale ratio $\epsilon = T_r/T_f$ alters the relative rapidity of the dynamic forcing. Furthermore, this input admits closed-form analytical solutions for both the exact system and the ODE model, enabling quantitative evaluation without numerical integration error.
\begin{figure}[t]
\centering
\includegraphics[width=\linewidth]{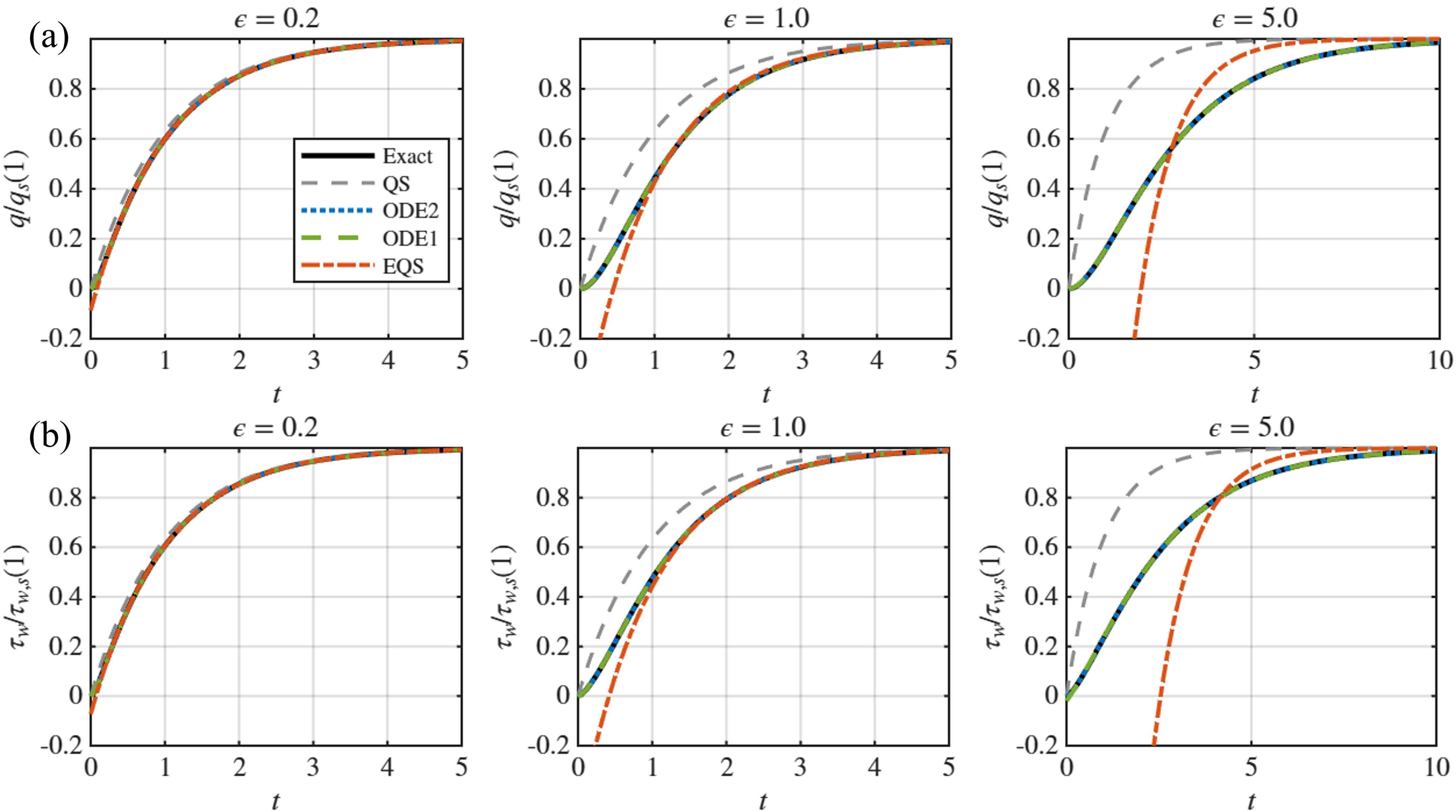}
\caption{Comparison of normalized transient responses under exponential ramp forcing for timescale ratios $\epsilon=0.2$, $1.0$, and $5.0$: (a) normalized volumetric flow rate $q/q_s(1)$ and (b) reconstructed wall shear stress $\tau_w/\tau_{w,s}(1)$. Results are compared among the exact solution, quasi-steady (QS) approximation, extended quasi-steady (EQS) approximation, and the ODE models with $p=1$(ODE1) and  $p=2$(ODE2).}
\label{fig:2}
\end{figure}

Figure~\ref{fig:2} compares the transient responses under exponential forcing across different values of $\epsilon$. The quasi-steady approximation neglects transient memory and therefore responds instantaneously to changes in the forcing. The EQS approximation improves the slow-forcing response ($\epsilon \ll 1$) but deteriorates under rapid forcing ($\epsilon \sim 1$), whereas the ODE models retain the dominant viscous relaxation mode and reproduce the exact flow-rate response over a broad range of $\epsilon$. The predictions of the first-order ($p=1$) and second-order ($p=2$) closures are virtually indistinguishable across all tested timescale ratios, confirming that the single fundamental relaxation mode captured by the first-order model already accounts for nearly the entire dynamic response under continuous forcing. However, the short-time kernels reveal a difference between the two closures: compared with the first-order model, the second-order model reduces the spurious nonzero initial values of the flow rate and the wall shear arising from rational truncation.

Under rapid forcing ($\epsilon \gg 1$), the first-order ODE model exhibits a slight decrease in accuracy for wall shear compared to the bulk flow rate, though this minor discrepancy remains barely noticeable in the trajectory plots. This behavior is attributed to the underlying modal structure: while the fundamental mode accounts for over $98.6\%$ of the flow-rate response ($a_{q,0} \approx 0.986$), it captures only about $81.1\%$ of the wall shear stress ($a_{\tau,0} \approx 0.811$). Governed directly by localized near-wall velocity gradients, wall shear stress is inherently more sensitive to rapid diffusion modes omitted by the single-pole model. Advancing to the second-order model ($p=2$) incorporates the second relaxation mode ($\kappa_2 \approx 23.15 \approx \gamma_1$), thereby capturing this secondary modal contribution and refining the prediction of $\tau_w$ under rapid transients.
%======================================================
\subsection{Nonlinear extension to turbulent flow}
We extend the framework to unsteady turbulent flows governed by a nonlinear Reynolds-averaged Navier-Stokes (RANS) model. Unlike laminar flow, the effective momentum diffusivity depends on the velocity gradient and therefore varies with the instantaneous flow state. This makes both the steady flow-rate relation and the local dynamic response state-dependent. The following analyses focus on the bulk flow rate $q$ using the $p=1$ closure. Because spatial integration suppresses higher diffusion modes—a mechanism further enhanced in turbulent flow by eddy-viscosity dissipation—the $p=1$ formulation is expected to capture the dominant dynamic memory of $q$ with reasonable accuracy.

Using the same nondimensionalization as in the preceding laminar analysis, the Reynolds-averaged momentum equation with Prandtl's mixing-length closure is
\begin{equation}
\epsilon \frac{\partial u}{\partial t} = F(t) + \frac{\partial}{\partial y} \left[ \left( 1 + Re_r l_m^2 \left| \frac{\partial u}{\partial y} \right| \right) \frac{\partial u}{\partial y} \right],
\label{eq:turbulent_pde}
\end{equation}
where $Re_r = q_r / \nu$ is the Reynolds number based on the representative dimensional flow rate per unit span, and $l_m(y) = \kappa(1-|y|)$ is the dimensionless mixing length with $\kappa = 0.41$. Equation~\eqref{eq:turbulent_pde} defines the full-order benchmark model.
For each constant value of $F$, the corresponding steady velocity profile $u_s(y;F)$ satisfies
\begin{equation}
0 = F + \frac{\mathrm{d}}{\mathrm{d}y} \left[ \left( 1 + Re_r l_m^2 \left| \frac{\mathrm{d}u_s}{\mathrm{d}y} \right| \right) \frac{\mathrm{d}u_s}{\mathrm{d}y} \right].
\label{eq:turbulent_steady_pde}
\end{equation}
The steady flow-rate family $q_s(F)$ and its differential response $q_s'(F) = \mathrm{d}q_s/\mathrm{d}F$ are evaluated numerically by integrating $u_s(y;F)$ over $y \in [-1, 1]$:
\begin{equation}
q_s(F) = \int_{-1}^{1} u_s(y;F) \, \mathrm{d}y.
\label{eq:turbulent_steady_manifold}
\end{equation}
The function $q_s(F)$ is interpolated directly along the family of steady states. 

Because $l_m=0$ at the walls, cross-channel integration recovers the same integral balance in Eq.~\eqref{eq:momentum} and $\mathcal{U} \equiv -2\Delta\tau_w = \epsilon\dot{q}$. The state-dependent closure of $p=1$ along the steady family is then expressed as
\begin{equation}
\mathcal{U} \simeq C_0(F)\Delta q + \epsilon D_1(F)\dot{F}.
\label{eq:turbulent_nonlinear_closure}
\end{equation}
Defining the model prediction as $q_{\mathrm{ode}}(t)\equiv q_s(F(t))+\Delta q(t)$, the state-dependent ODE becomes
\begin{equation}
\epsilon \dot{q}_{\mathrm{ode}}=C_0(F)\left[q_{\mathrm{ode}}-q_s(F)\right]+\epsilon D_1(F)\dot{F}.
\label{eq:turbulent_nonlinear_ode}
\end{equation}
Linearizing this model about a fixed operating state $\bar{F}$, where $\delta F=F-\bar{F}$ and $\delta q_{\mathrm{ode}}=q_{\mathrm{ode}}-q_s(\bar{F})$ denote small perturbations of the forcing and flow rate from their steady-state values, yields the local flow-rate transfer function:
\begin{equation}
G^\mathrm{loc}_q(\sigma;\bar{F}) \equiv \frac{\widehat{\delta q}_{\mathrm{ode}}(\sigma)}{q_s'(\bar{F})\widehat{\delta F}(\sigma)}
=1+\frac{\left[D_1(\bar{F})/q_s'(\bar{F})-1\right]\sigma}{\sigma-C_0(\bar{F})},
\label{eq:turbulent_local_transfer}
\end{equation}
which satisfies $G^\mathrm{loc}_q(0;\bar{F})=1$. Here, we adopt low-frequency moment matching as one possible coefficient-identification strategy. The state-dependent closure coefficients $C_0(\bar F)$ and $D_1(\bar F)$ are determined offline along the family of steady states by matching the low-frequency expansion of $G^\mathrm{loc}_q(\sigma;\bar{F})$ to the corresponding moments of the linearized PDE resolvent associated with Eq.~\eqref{eq:turbulent_pde} (see Appendix~A). The resulting coefficient functions are then smoothly interpolated over the operating range of $F$ and used in the evaluation of Eq.~\eqref{eq:turbulent_nonlinear_ode}.

To evaluate the predictive capability of the state-dependent ODE model, we consider acceleration (Acc) and deceleration (Dec) transients between two steady states $(F_0,q_0)$ and $(F_1,q_1)$, where $q_0=q_s(F_0)$ and $q_1=q_s(F_1)$. The imposed forcing $F(t)$ and normalized response progress $q_N(t)$ are defined as
\begin{equation}
F(t) =
\begin{cases}
F_0 + (F_1 - F_0)(1 - e^{-t}), & \mathrm{Acc},\\
F_1 + (F_0 - F_1)(1 - e^{-t}), & \mathrm{Dec},
\end{cases}
\qquad
q_N(t) =
\begin{cases}
\dfrac{q(t) - q_0}{q_1 - q_0}, & \mathrm{Acc},\\[1.2ex]
\dfrac{q_1 - q(t)}{q_1 - q_0}, & \mathrm{Dec}.
\end{cases}
\label{eq:forcing_and_response}
\end{equation}

For the full-order RANS benchmark, Eq.~\eqref{eq:turbulent_pde} is discretized in space using a cell-centered finite-volume scheme on a cosine-clustered grid with $N_y = 200$ cells, as determined by a grid-sensitivity analysis. The resulting system of coupled ODEs is integrated in time using MATLAB's ode15s solver with relative tolerances of $2 \times 10^{-8}$. To construct the family of steady states, steady profiles $u_s(y;F)$ and the corresponding linearized Jacobians $\boldsymbol{J}(F)$ are evaluated at 26 anchor forcing levels over $F \in [F_0, 10F_0]$. We set $Re_r=3000$; the reference forcing $F_0\simeq10.62$ is chosen such that $q_s(F_0)=1$, giving a friction Reynolds number $Re_\tau = \sqrt{Re_r F_0} \approx 178.5$. At each anchor state, the local ODE coefficients $C_0(F)$ and $D_1(F)$ are obtained by matching the low-frequency moments of the linearized flow-rate transfer function. These coefficients, together with the steady-state quantities, are interpolated along the family of steady states using piecewise cubic Hermite interpolation. Finally, the ODE is integrated using MATLAB's ode45 solver with relative tolerances of $10^{-8}$.

Across the forcing range $F/F_0 \in [1, 10]$, the system dynamics are dominated by two main state-dependent terms: the steady flow rate $q_s(F)$ increases sublinearly from $1$ to $4.2$, and the relaxation coefficient $-C_0(F)$ increases from approximately $32$ to $80$, indicating a $2.5$-fold acceleration of the local dynamic relaxation as momentum diffusion intensifies at higher forcing levels. This is approximately one order of magnitude larger than the laminar relaxation coefficient. The forcing-rate coefficient $D_1(F)$ remains small relative to the steady-state sensitivity, with $D_1(F)/q_s'(F)=\mathcal{O}(10^{-3})$ (decreasing from $1.1\times10^{-3}$ to $0.7\times10^{-3}$). Consequently, the transient flow-rate response is governed primarily by $q_s(F)$ and $C_0(F)$, with $D_1(F)$ providing only a minor correction.

We evaluate 12 finite-amplitude transients, consisting of acceleration and deceleration cases for two forcing amplitude ratios, $F_1/F_0=2.0$ and $10.0$, and three timescale ratios, $\epsilon_0 \in \{0.2,1.0,5.0\}$. To characterize the forcing rate relative to the relaxation dynamics, we define $\epsilon_0=T_{0}(F_0)/T_f$, where $T_{0}(F)=T_r/[-C_0(F)]$ is the local relaxation time. The same reference state $F_0$ is used to define $\epsilon_0$ for both acceleration and deceleration, so that each pair is compared at the same forcing timescale $T_f$. Since the governing equations use $\epsilon=T_r/T_f$, the corresponding parameter is $\epsilon=-C_0(F_0)\epsilon_0$. Because $C_0(F)$ varies with the operating state, the local ratio of the relaxation time to the forcing timescale also varies during the transient as $\epsilon/[-C_0(F)]$. For comparison, we also evaluate the quasi-steady (QS) approximation and an initial-state-frozen ODE, in which $q_s(F)$ varies with the instantaneous forcing while $C_0$ and $D_1$ are fixed at their initial steady-state values.

\begin{figure}[t]
\centering
\includegraphics[width=\textwidth]{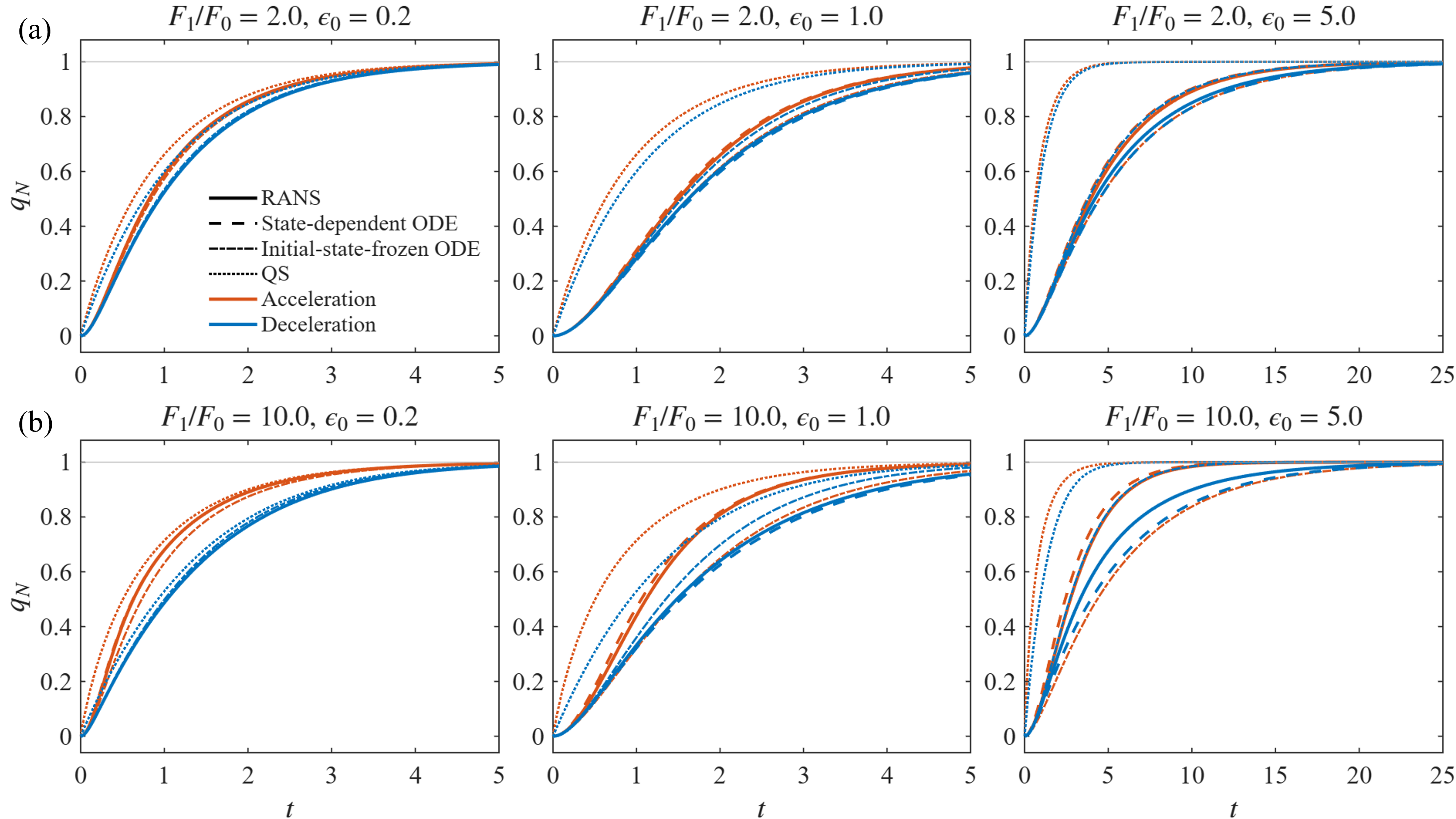}
\caption{Comparison of normalized transient flow-rate responses $q_N(t)$ under finite-amplitude exponential forcing for acceleration (red) and deceleration (blue). Columns correspond to timescale ratios $\epsilon_0 = 0.2$, $1.0$, and $5.0$, while rows correspond to forcing-amplitude ratios $F_1/F_0=2$ and $10$. The curves show the full-order PDE solution (solid), the state-dependent ODE (dashed), the initial-state-frozen ODE (dash-dotted), and the quasi-steady approximation (dotted).}
\label{fig:3}
\end{figure}

Figure~\ref{fig:3} compares the transient flow-rate responses across all test conditions. An acceleration–deceleration asymmetry is present even in the QS predictions due to the nonlinear steady-state relation $q_s(F)$. The full-order responses additionally exhibit state-dependent relaxation, with the local relaxation rate $-C_0(F)$ increasing as turbulent mixing intensifies at higher forcing levels. By tracking $C_0(F)$ along the family of steady states, the state-dependent ODE captures the resulting dynamic departures from the QS response. For both forcing amplitude ratios, $F_1/F_0 = 2.0$ and $10.0$, the state-dependent ODE agrees with the RANS benchmarks for all $\epsilon_0 \le 1.0$. Significant discrepancies appear only under the most rapid, large-amplitude forcing ($F_1/F_0=10.0$ and $\epsilon_0=5.0$): the ODE slightly leads the RANS response during acceleration, whereas it lags during deceleration. These discrepancies may reflect departures of the transient flow profile from the local steady-profile family during rapid forcing.

The initial-state-frozen model remains reasonable only under small-amplitude, slow forcing ($F_1/F_0=2.0$ and $\epsilon_0=0.2$) and deteriorates significantly as either the amplitude ratio or $\epsilon_0$ increases. Notably, at larger $\epsilon_0$, a crossover phenomenon emerges: the frozen acceleration solution (fixed at the initial rate $-C_0(F_0)$) aligns better with the late-stage deceleration trajectory, whereas the frozen deceleration solution (fixed at $-C_0(F_1)$) aligns with the late-stage acceleration trajectory. This crossover occurs because the true flow dynamics ultimately relax at the rate corresponding to the final state rather than the initial state, demonstrating that a single fixed relaxation pole cannot capture the evolving local dynamics. 

Overall, combining the nonlinear steady-state family $q_s(F)$ with state-dependent relaxation dynamics enables accurate prediction of finite-amplitude transients and directional asymmetry without fitting the ODE model to unsteady training trajectories. Although a state-dependent second-order closure was also examined (results not shown), it yielded only limited improvement in predictive accuracy, suggesting that the first-order closure captures the dominant relaxation dynamics. The remaining discrepancies under rapid, large-amplitude forcing may therefore reflect nonlinear dynamic effects not represented by the locally linearized closure.
%===========================================================================================================
\section{Canonical Validation II: Wave-Propagation Dynamics}
%===========================================================================================================
This section considers Saint--Venant flow in a finite-length channel, where wave propagation and reflection introduce spatial memory that is not directly resolved by the integral momentum balance. The dynamic response of shallow open-channel flows has been studied using linear wave approximations \cite{Ponce1977} and low-order transfer-function models developed for flow control \cite{Litrico2004,Munier2008}. Low-order nonlinear reduced models based on local linearization and low-frequency matching have also been developed \cite{Litrico2010}. Here, we construct the reduced dynamics from the integral momentum balance and represent the remaining unresolved spatial effects using the temporal closure framework introduced in Section~II.

\subsection{Governing system, steady-state family, and exact momentum balance}
We consider one-dimensional shallow-water wave dynamics in a uniform rectangular open channel of width $b$ and length $l$. The system is linearized about a quiescent reference state of uniform depth $h_r$ and zero discharge, while transient flow resistance is represented by an equivalent linear friction coefficient $f$. The quiescent reference state is chosen to eliminate background-flow effects, allowing the wave-propagation and frictional-dissipation dynamics to be examined in isolation.

The gravity-wave celerity and wave-propagation timescale are defined as $c=\sqrt{gh_r}$ and $T_r=l/c$, respectively. With a characteristic forcing timescale $T_f$, the dimensionless parameters $\epsilon$ and $\mu$ are defined as
\begin{equation}
\epsilon \equiv \frac{T_r}{T_f} = \frac{l}{\sqrt{g h_r} T_f}, \qquad \mu \equiv f T_r = \frac{f l}{\sqrt{g h_r}}.
\end{equation}
Here, $\epsilon$ represents the ratio of the internal wave-propagation delay to the external forcing timescale, while $\mu$ measures the relative strength of frictional dissipation over a wave-propagation timescale.

Space and time are nondimensionalized as $x=\tilde{x}/l\in[0,1]$ and $t=\tilde{t}/T_f$. The free-surface elevation perturbation $\tilde{\eta}$ is scaled by $h_r$, while the local discharge perturbation $\tilde{\mathcal{Q}}$ is scaled by the characteristic discharge $Q_r=bh_rc/\mu$. The resulting dimensionless variables are denoted by $\eta(x,t)$ and $\mathcal{Q}(x,t)$.

The prescribed upstream boundary-head perturbation $H(t)\equiv\eta(0,t)$ serves as the forcing, while the downstream discharge $Q(t)\equiv\mathcal{Q}(1,t)$ is the bulk response of interest. Under these scalings, the dimensionless linear Saint--Venant system is
\begin{equation}
\epsilon\mu\frac{\partial\eta}{\partial t}+\frac{\partial\mathcal{Q}}{\partial x}=0,\qquad \epsilon\frac{\partial\mathcal{Q}}{\partial t}+\mu\frac{\partial\eta}{\partial x}+\mu\mathcal{Q}=0.
\label{eq:SV_pde}
\end{equation}
The boundary conditions are $\eta(0,t)=H(t)$ and $\eta(1,t)=0$. The fixed-stage downstream boundary defines a surface-elevation node for wave reflection and, through the continuity equation, implies $\partial\mathcal{Q}/\partial x=0$ at $x=1$. The downstream discharge is therefore free to vary dynamically while the downstream stage remains fixed.

For constant forcing $H$, the steady solution is
\begin{equation}
\eta_s(x;H)=H(1-x),\qquad Q_s(H)=H,
\label{eq:SV_steady}
\end{equation}
which defines a one-parameter steady-state family with $Q_s'(H)\equiv\mathrm{d}Q_s/\mathrm{d}H=1$. Integrating the momentum equation in Eq.~\eqref{eq:SV_pde} over $x\in[0,1]$ gives the exact integral momentum balance
\begin{equation}
\epsilon\dot M= -\mu M+ \mu H(t),\qquad M(t)\equiv\int_0^1\mathcal{Q}(x,t)\,\mathrm{d}x.
\label{eq:SV_momentum}
\end{equation}
In the notation of Section~II, $Y=Q$, $\lambda=H$, and $I=M$. Since $M_s(H)=Q_s(H)=H$, the departures are
\begin{equation}
\Delta M=M-H,\qquad \Delta Q=Q-H.
\label{eq:SV_departures}
\end{equation}
The exact balance then becomes
\begin{equation}
\epsilon\dot H+\epsilon\dot{\Delta M}=-\mu\Delta M.
\label{eq:SV_momentum_departure}
\end{equation}
Although this balance governs the integral storage $M$ exactly, it does not directly determine the resolved downstream discharge $Q$. A temporal closure is therefore required to account for the difference $M-Q$.
%======================================================
\subsection{Spatial structure and low-order dynamic closure}
The integral storage departure is decomposed as $\Delta M=\Delta Q+(M-Q)$, where $\Delta Q=Q-H$ is the resolved discharge departure and $M-Q$ is the unresolved storage contribution. Substitution into the exact integral balance gives
\begin{equation}
\epsilon\dot H+\epsilon\dot{\Delta Q}+\mu\Delta Q=\mathcal U,\qquad
\mathcal U=-(\mu+\epsilon \frac{d}{dt})(M-Q).
\end{equation}
The continuity equation in Eq.~\eqref{eq:SV_pde} provides an exact relation for this difference. Multiplying that equation by $x$ and integrating over the channel yields
\begin{equation}
M-Q=\epsilon\mu\frac{\mathrm d}{\mathrm dt}\int_0^1x\eta(x,t)\,\mathrm dx.
\label{eq:mq_closure}
\end{equation}
To examine the temporal derivative structure induced by the spatial dynamics, we eliminate $\eta$ from Eq.~\eqref{eq:SV_pde} to obtain the telegraph equation
\begin{equation}
\frac{\partial^2\mathcal{Q}}{\partial x^2}=\epsilon^2\frac{\partial^2\mathcal{Q}}{\partial t^2}+\epsilon\mu\frac{\partial\mathcal{Q}}{\partial t}.
\label{eq:Qequation}
\end{equation}
Expanding $\mathcal{Q}(x,t)$ about the downstream boundary $x=1$, using $\mathcal{Q}(1,t)=Q(t)$ and $\left.\partial\mathcal{Q}/\partial x\right|_{x=1}=0$, gives
\begin{equation}
\begin{aligned}
\mathcal{Q}(x,t)
&=Q(t)
+\frac{(x-1)^2}{2}
\left[\epsilon^2\ddot Q(t)+\epsilon\mu\dot Q(t)\right] \\
&\quad+\frac{(x-1)^4}{24}
\left[
\epsilon^4 Q^{(4)}(t)
+2\epsilon^3\mu Q^{(3)}(t)
+\epsilon^2\mu^2\ddot Q(t)
\right]+\mathcal{O}\!\left((x-1)^6\right).
\end{aligned}
\label{eq:Texpansion}
\end{equation}
All odd-order spatial derivatives vanish at $x=1$, as follows from Eq.~\eqref{eq:Qequation} and the downstream boundary condition. The expansion illustrates how spatial corrections introduce temporal derivatives into the integral momentum storage. Since the exact momentum balance also involves the time derivative of this storage, higher-order temporal derivatives arise naturally in the reduced dynamics. In the present study, we consider temporal closure orders up to $p=5$.

The exact spatial-moment relation in Eq.~\eqref{eq:mq_closure} motivates a derivative-only approximation of the unresolved contribution, corresponding to $C_{0,p}=0$ in the general closure of Eq.~\eqref{eq:U_general_nondim}. For the reduced models considered here, we adopt
\begin{equation}
\mathcal U \simeq \sum\limits_{k = 1}^p {\left[ {\epsilon^k C_{k,p} \Delta Q^{(k)}  + \epsilon^k D_{k,p} H^{(k)} } \right]} ,\qquad C_{0,p}  = 0,\quad p \le 5.
\label{eq:SV_closure_p3_general}
\end{equation}

Substituting this closure into the exact departure balance gives the $p$th-order reduced dynamics
\begin{equation}
\sum\limits_{k = 0}^p {\epsilon^k A_{k,p} \Delta Q^{(k)} }  =  - \sum\limits_{k = 1}^p {\epsilon^k B_{k,p} H^{(k)} } 
\label{eq:SV_departure_general}
\end{equation}
where the composite coefficients are defined as

\begin{equation}
\begin{aligned}
&A_{0,p}=1,\quad
A_{1,p}=\frac{1-C_{1,p}}{\mu},\quad
B_{1,p}=\frac{1-D_{1,p}}{\mu},\\
&A_{k,p}=-\frac{C_{k,p}}{\mu},\quad
B_{k,p}=-\frac{D_{k,p}}{\mu},
\qquad k=2,\ldots,p.
\end{aligned}
\label{eq:SV_AB_general}
\end{equation}
Once the closure coefficients $C_{k,p}$ and $D_{k,p}$ are specified, the corresponding ODE model and its EQS expansion can be constructed for any order $p\le5$.
%============================================
\subsection{Linear response and coefficient identification}
We next identify the closure coefficients from the frequency response of the linear system. We introduce the intrinsic-time-scaled Laplace variable $\sigma=\epsilon s$, which reduces to $\sigma=i\epsilon\omega$ for harmonic forcing at the dimensionless angular frequency $\omega=\tilde{\omega}T_f$. For zero initial perturbations, the inverse of the normalized exact transfer function is
\begin{equation}
G_Q(\sigma)^{-1} \equiv \frac{Q_s'\hat H(\sigma)}{\hat Q(\sigma)} = \left(1+\frac{\sigma}{\mu}\right) \frac{\sinh z}{z}, \qquad  z=\sqrt{\sigma^2+\mu \sigma}.
\label{eq:exact_transfer_function_inv}
\end{equation}
Since $z^2=O(\sigma)$, the factor $\sinh z/z$ is expanded through fifth order in $\sigma$ as
\begin{equation}
\frac{\sinh z}{z} =\sum_{k=0}^{5}\frac{z^{2k}}{(2k+1)!}+O(\sigma^6).
\label{eq:spatial_operator_truncation}
\end{equation}
Accordingly, the exact transfer function has the low-frequency expansion
\begin{equation}
G_Q(\sigma)=1+\sum_{k=1}^{5}g_k \sigma^k+O(\sigma^6),
\label{eq:SV_transfer_leading_curvature}
\end{equation}
where
\begin{equation}
\begin{aligned}
g_1&=-\frac{1}{\mu}-\frac{\mu}{6}, \quad g_2=\frac{1}{\mu^2}+\frac{7\mu^2}{360}, \quad g_3=-\frac{1}{\mu^3}+\frac{7\mu}{360}-\frac{31\mu^3}{15120}, \\
g_4&=\frac{1}{\mu^4}-\frac{31\mu^2}{7560}+\frac{127\mu^4}{604800}, \quad g_5=-\frac{1}{\mu^5}-\frac{31\mu}{15120}+\frac{127\mu^3}{201600}-\frac{73\mu^5}{3421440}.
\end{aligned}
\end{equation}
Taking the Laplace transform of Eq.~\eqref{eq:SV_departure_general} for zero initial perturbations and using $\Delta Q=Q-H$ gives the general
reduced transfer function
\begin{equation}
G^{\mathrm{ode}}_{Q,p} (\sigma) \equiv \frac{\hat Q(\sigma)}{{Q'_s \hat H(\sigma)}} = \frac{{1 + \sum\limits_{k = 1}^p ( A_{k,p}  - B_{k,p} )\sigma^k }}{{1 + \sum\limits_{k = 1}^p {A_{k,p} } \sigma^k }}.
\label{eq:SV_rom_transfer_general}
\end{equation}
The denominator polynomials are constructed by retaining the common relaxation pole $\sigma=-\mu$ given by Eq.~\eqref{eq:exact_transfer_function_inv} and matching the low-frequency behavior of the remaining spatial dynamics, yielding the expressions summarized in Table~\ref{tab:SV_denominator_coefficients}. The first-order model ($p=1$) contains only this relaxation pole. At $p=2$, an additional real pole approximates the leading low-frequency contribution of the wave modes. The third-order model ($p=3$) explicitly incorporates the first wave-mode pair through a quadratic factor. The fourth-order model ($p=4$) retains this pair and adds a real pole to approximate the contribution of the remaining modes, while the fifth-order model ($p=5$) further incorporates the second wave-mode pair. 

The first and second wave-mode pairs are characterized by the poles
\begin{equation}
\sigma_1^\pm=\frac{-\mu\pm\sqrt{\mu^2-4\pi^2}}{2}, \qquad \sigma_2^\pm=\frac{-\mu\pm\sqrt{\mu^2-16\pi^2}}{2}.
\label{eq:wavepoles}
\end{equation}
The first and second pole pairs correspond to damped oscillatory modes when $\mu<2\pi$ and $\mu<4\pi$, respectively; otherwise, the corresponding poles are real. The third- and fourth-order models incorporate the first wave-mode pair, while the fifth-order model additionally incorporates the second.

\begin{table*}[t]
\centering
\caption{Structured denominator polynomials and their coefficients for the linear Saint--Venant reduced models.}
\label{tab:SV_denominator_coefficients}
\begingroup
\renewcommand{\arraystretch}{1.8}
\setlength{\tabcolsep}{6pt}
\resizebox{\textwidth}{!}{%
\begin{tabular}{@{}c l c c c c c@{}}
\hline
$p$ & $1+\displaystyle\sum_{k=1}^{p} A_{k,p}\sigma^k$
& $A_{1,p}$ & $A_{2,p}$ & $A_{3,p}$ & $A_{4,p}$ & $A_{5,p}$ \\
\hline
1
& $\displaystyle 1+\frac{\sigma}{\mu}$
& $\displaystyle \frac{1}{\mu}$
& -- & -- & -- & -- \\
2
& $\displaystyle \left(1+\frac{\sigma}{\mu}\right)\left(1+\frac{\mu \sigma}{6}\right)$
& $\displaystyle \frac{1}{\mu}+\frac{\mu}{6}$
& $\displaystyle \frac{1}{6}$
& -- & -- & -- \\
3
& $\displaystyle \left(1+\frac{\sigma}{\mu}\right)
\left(1+\frac{\sigma^2+\mu \sigma}{\pi^2}\right)$
& $\displaystyle \frac{1}{\mu}+\frac{\mu}{\pi^2}$
& $\displaystyle \frac{2}{\pi^2}$
& $\displaystyle \frac{1}{\mu\pi^2}$
& -- & -- \\
4
& $\displaystyle \left(1+\frac{\sigma}{\mu}\right)
\left(1+\frac{\sigma^2+\mu \sigma}{\pi^2}\right)
\left[1+\left(\frac{1}{6}-\frac{1}{\pi^2}\right)\mu\sigma\right]$
& $\displaystyle \frac{1}{\mu}+\frac{\mu}{6}$
& $\displaystyle \frac{1}{6}+\frac{1}{\pi^2}
+\frac{\mu^2}{\pi^2}\left(\frac{1}{6}-\frac{1}{\pi^2}\right)$
& $\displaystyle \frac{1}{\mu\pi^2}
+\frac{2\mu}{\pi^2}\left(\frac{1}{6}-\frac{1}{\pi^2}\right)$
& $\displaystyle \frac{1}{\pi^2}\left(\frac{1}{6}-\frac{1}{\pi^2}\right)$
& -- \\
5
& $\displaystyle \left(1+\frac{\sigma}{\mu}\right)
\left(1+\frac{\sigma^2+\mu \sigma}{\pi^2}\right)
\left(1+\frac{\sigma^2+\mu \sigma}{4\pi^2}\right)$
& $\displaystyle \frac{1}{\mu}+\frac{5\mu}{4\pi^2}$
& $\displaystyle \frac{5}{2\pi^2}+\frac{\mu^2}{4\pi^4}$
& $\displaystyle \frac{5}{4\mu\pi^2}+\frac{3\mu}{4\pi^4}$
& $\displaystyle \frac{3}{4\pi^4}$
& $\displaystyle \frac{1}{4\mu\pi^4}$ \\
\hline
\end{tabular}%
}
\endgroup
\end{table*}
Once the denominator coefficients $A_{k,p}$ are specified, the remaining coefficients $B_{k,p}$ follow by matching the low-frequency expansion of Eq.~\eqref{eq:SV_rom_transfer_general} to the exact series in Eq.~\eqref{eq:SV_transfer_leading_curvature}:
\begin{equation}
B_{k,p}=-g_k-\sum_{j=1}^{k-1}A_{j,p}g_{k-j}, \qquad k = 1, \ldots ,p,
\label{eq:SV_numerator_coefficients}
\end{equation}
Thus, the same coefficient-identification procedure applies to all orders considered, and each reduced transfer function matches the exact low-frequency expansion through order $\sigma^p$.

Substituting the identified composite coefficients into Eq.~\eqref{eq:SV_departure_general} gives the total-response ODE for any $p=1,\ldots,5$:
\begin{equation}
Q_{\mathrm{ode},p}(t)+\sum_{k=1}^{p}\epsilon^kA_{k,p}Q_{\mathrm{ode},p}^{(k)}(t) =H(t)+\sum_{k=1}^{p}\epsilon^k  (A_{k,p}-B_{k,p})H^{(k)}(t).
\label{eq:SV_ode_general_identified}
\end{equation}
Alternatively, an extended quasi-steady (EQS) solution follows by expanding the exact low-frequency response:
\begin{equation}
Q_{\mathrm{eqs},p}(t)=H(t)+\sum_{k=1}^{p}\epsilon^kg_kH^{(k)}(t)+O(\epsilon^{p+1}),\qquad p=1,\ldots,5.
\label{eq:SV_EQS_general_identified}
\end{equation}

\begin{figure}[t]
\centering
\includegraphics[width=\linewidth]{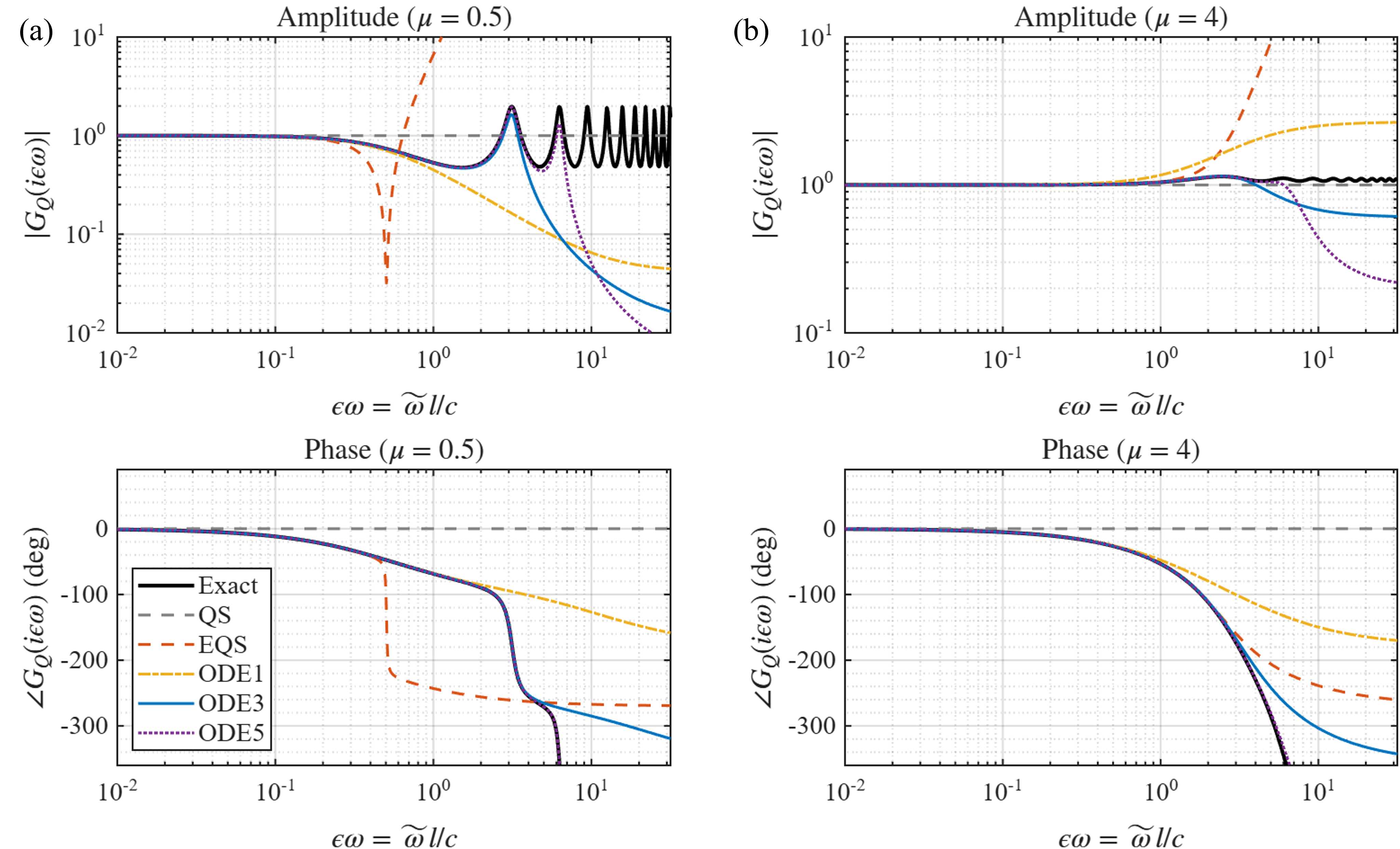}
\caption{Comparison of the transfer-function magnitude $|G_Q(i\epsilon \omega)|$ (top) and phase $\angle G_Q(i\epsilon \omega)$ (bottom) under harmonic forcing for two friction regimes: (a) $\mu=0.5$ (inertia-dominated) and (b) $\mu=4.0$ (friction-dominated). The exact response is compared against QS, EQS($p=5$), and the ODE hierarchy ($p=1,3,5$).}
\label{fig:4}
\end{figure}

Figure~\ref{fig:4} compares the exact linear transfer function with the QS, EQS ($p=5$), and ODE representations ($p=1,3,5$) under inertia-dominated ($\mu=0.5$) and friction-dominated ($\mu=4.0$) regimes. In the exact response, the inertia-dominated regime exhibits pronounced resonant behavior, characterized by a series of peaks associated with multiple wave reflections at the channel boundaries. In contrast, increased friction in the friction-dominated regime suppresses these oscillations and substantially attenuates the resonant peaks.

The algebraic approximations exhibit clear limitations. The standard QS model is restricted to the low-frequency limit and cannot capture any phase lag. Although EQS extends the range of validity and reproduces the initial phase lag, its polynomial dependence on frequency leads to an unbounded magnitude response at higher frequencies.

The first-order ODE model represents only relaxation dynamics and therefore cannot capture the resonant behavior associated with wave propagation. In contrast, the higher-order models ($p=3,5$) reproduce the oscillatory response. The third-order model captures the first resonance in both magnitude and phase, while the fifth-order model additionally incorporates the second wave-mode pair, extending the frequency range over which the exact response is accurately reproduced. The improvement is particularly evident in the inertia-dominated regime, where resonant effects are more pronounced.

\subsection{Response to transient forcing}
For zero initial perturbations and $H(0)=0$, the exact downstream discharge response can be written as a convolution with the kernel $K_Q(t)$:

\begin{equation}
Q(t)=\int_0^tK_Q(t-\xi)\dot H(\xi)\,\mathrm{d}\xi.
\label{eq:conv_exact_sv}
\end{equation}
The exact kernel is derived via the inverse Laplace transform of $G_Q(\epsilon s)/s$ as
\begin{equation}
K_Q(t)=1-e^{-\mu t/\epsilon}+2\mu\sum_{n=1}^{\infty} \frac{(-1)^n}{\varpi_n} e^{-\mu t/(2\epsilon)} \sin\left(\frac{\varpi_n}{\epsilon}t\right),
\label{eq:kernel_exact_sv}
\end{equation}
where $\varpi_n=\sqrt{n^2\pi^2-\mu^2/4}$. Equation~\eqref{eq:kernel_exact_sv} decomposes the exact response to a step change in the upstream head into a steady-state contribution, a non-oscillatory global relaxation mode $-e^{-\mu t/\epsilon}$, and an infinite hierarchy of damped wave modes governing propagation, attenuation, and boundary reflection \citep{Shimozono2021}. These modal terms are oscillatory when $\varpi_n$ is real, but become overdamped as friction increases. In this sense, Eq.~\eqref{eq:kernel_exact_sv} provides a finite-length, wave-dynamic counterpart to the classical diffusion kernel \citep{Hayami1951}, while explicitly retaining downstream reflections and the resulting infinite-dimensional spatial memory. 

The ODE models replace the infinite modal hierarchy by a finite number of temporal modes. The corresponding downstream response retains the convolution form of Eq.~\eqref{eq:conv_exact_sv}, with the exact kernel $K_Q$ replaced by the reduced kernel $K^{\mathrm{ode}}_{Q,p}$.

For $p=1,\ldots,5$, the reduced kernel can be expressed in the unified modal form

\begin{equation}
K^{\mathrm{ode}}_{Q,p}(t)
=
1-\sum_{j=1}^{p}
\frac{
\displaystyle\sum_{k=1}^{p}
B_{k,p} \sigma_{j,p}^{k-1}
}{
\displaystyle\sum_{k=1}^{p}
kA_{k,p} \sigma_{j,p}^{k-1}
}
e^{\sigma_{j,p}t/\epsilon},
\label{eq:kernel_rom_general}
\end{equation}
where $\sigma_{j,p}$ ($j=1,\ldots,p$) are the roots of the denominator of $G^{\mathrm{ode}}_{Q,p}$, ordered as

\begin{equation}
\begin{aligned}
\sigma_{1,p} &=-\mu,
\qquad p=1,\ldots,5,\\
\sigma_{2,2}&=-\frac{6}{\mu},\\
(\sigma_{2,p},\sigma_{3,p})&=(\sigma_1^+,\sigma_1^-),\qquad p=3,4,5,\\
\sigma_{4,4}&=-\frac{6\pi^2}{\mu(\pi^2-6)},\\
(\sigma_{4,5},\sigma_{5,5})&=(\sigma_2^+,\sigma_2^-),
\end{aligned}
\label{eq:rom_pole_ordering}
\end{equation}
where $\sigma_1^\pm$ and  $\sigma_2^\pm$ are given in Eq.~\eqref{eq:wavepoles}. The ordering is introduced solely for notational convenience and does not affect the kernel. Equation~\eqref{eq:kernel_rom_general} applies to distinct roots; coincident roots are handled by taking the appropriate limit.

Increasing $p$ enriches the modal structure of the kernel, allowing the reduced models to capture progressively more of the temporal dynamics associated with wave propagation and reflection. However, low-frequency moment matching does not guarantee accurate high-frequency behavior. Although the exact kernel satisfies $K_Q(0^+)=0$, the reduced kernels may be nonzero at $t=0^+$, allowing nonphysical downstream responses before the wave arrival time of the exact system.

The transient hierarchy is evaluated using the same exponential forcing as in the preceding benchmark, with the substitution $F\rightarrow H$ in Eq.~\eqref{eq:exp_ramp}. The prescribed upstream head and fixed downstream stage establish reflecting boundary conditions, allowing waves to undergo repeated reflections between the channel ends while being attenuated by friction. These reflections give rise to a discrete spectrum of damped wave modes. Rapid forcing excites a broad range of these modes, making this configuration a challenging benchmark. 

\begin{figure}[t]
\centering
\includegraphics[width=1\linewidth]{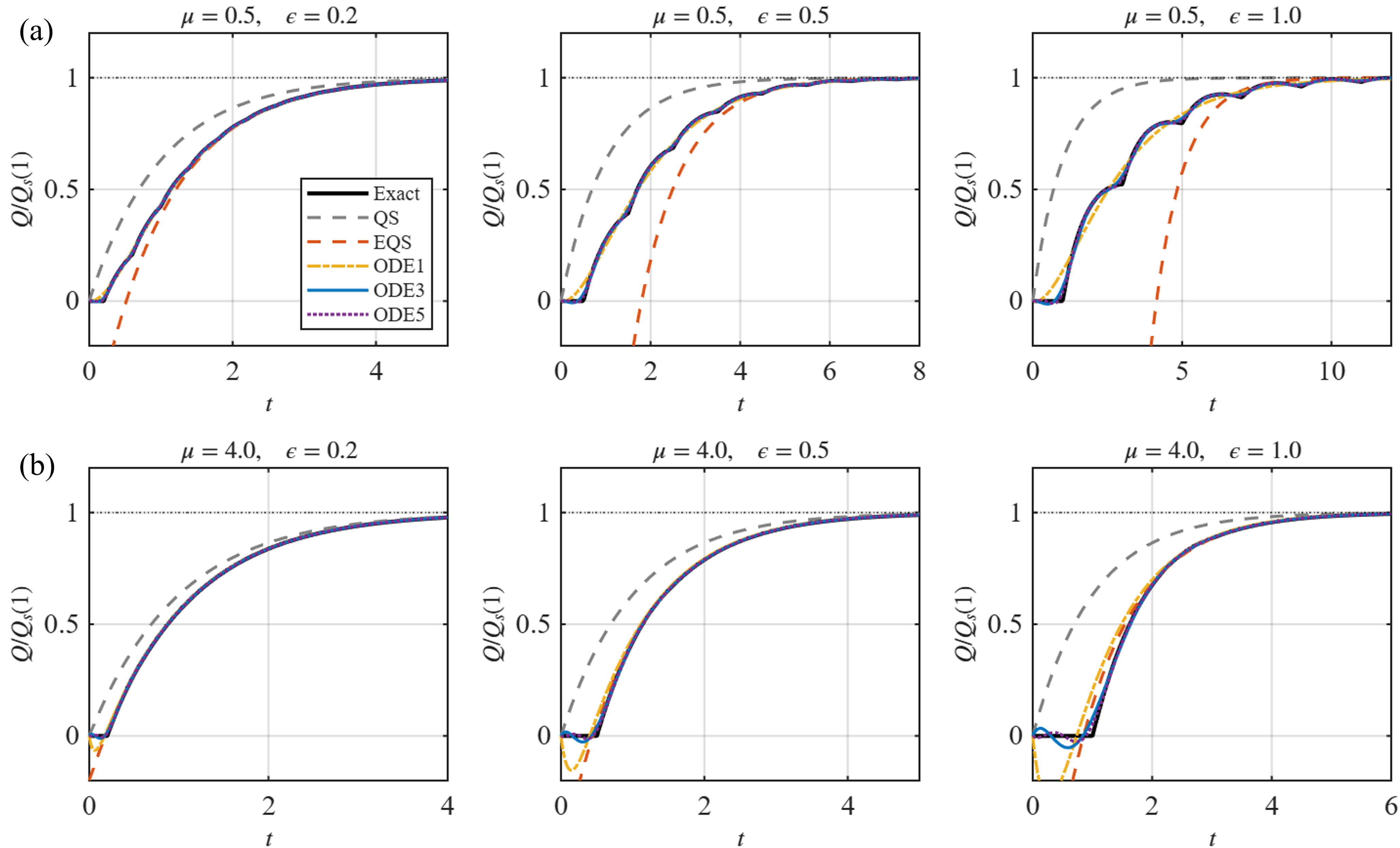}
\caption{Comparison of dimensionless transient responses under exponential ramp forcing for timescale ratios $\epsilon=0.2$, $0.5$, and $1.0$: (a) inertia-dominated regime ($\mu=0.5$) and (b) friction-dominated regime ($\mu=4.0$). The exact distributed solution is compared against QS, EQS ($p=5$), and the ODE hierarchy ($p=1,3,5$).}
\label{fig:5}
\end{figure}
Figure~\ref{fig:5} compares the normalized discharge responses $Q(t)/Q_s(1)$ across different timescale ratios ($\epsilon=0.2$, $0.5$, and $1.0$) to elucidate the performance of each model reduction scheme. In the low-friction regime ($\mu=0.5$), rapid forcing (large $\epsilon$) excites wave propagation and domain reflections, generating pronounced oscillatory transients. Under these conditions, the ODE hierarchy provides a closer approximation to the oscillatory response than the algebraic approximations. In particular, the higher-order ODE models ($p=3,5$) capture the dominant oscillatory memory through their wave-mode pairs, reproducing the transient envelope and principal phase delay. The fifth-order model further improves the transient response by incorporating the second wave-mode pair. In contrast, the EQS approximation exhibits substantial deviations under rapid forcing because its polynomial low-frequency expansion is applied outside its range of validity.

In the higher-friction regime ($\mu=4.0$), wave reflections are strongly attenuated, shifting the dynamics toward a relaxation-dominated response. Under rapid forcing ($\epsilon=1.0$), the exact solution exhibits an initial propagation delay followed by a relatively sharp rise, which the finite-order ODE models cannot reproduce exactly. Because low-frequency matching does not constrain the short-time response, the reduced kernels can exhibit a nonphysical initial undershoot. This artifact reflects the limited high-frequency accuracy of the finite-order approximation rather than an instability of the reduced dynamics. Nevertheless, increasing the model order improves the subsequent transient response, with the fifth-order model providing the closest approximation.
%======================================================
\subsection{Extension to nonlinear shallow-water dynamics}
We extend the ODE model to nonlinear Saint--Venant dynamics:
\begin{equation}
\epsilon\mu\frac{\partial\eta}{\partial t}+\frac{\partial\mathcal{Q}}{\partial x}=0,\qquad
\epsilon\frac{\partial\mathcal{Q}}{\partial t}
+\frac{1}{\mu}\frac{\partial}{\partial x}\left(\frac{\mathcal{Q}^2}{1+\eta}\right)
+\mu(1+\eta)\frac{\partial\eta}{\partial x}
+\mu\mathcal{Q}=0,
\label{eq:sv_pde_nonlinear}
\end{equation}
subject to the same boundary conditions $\eta(0,t)=H(t)$ and $\eta(1,t)=0$.

Integrating the nonlinear momentum equation in Eq.~\eqref{eq:sv_pde_nonlinear} over the channel yields the exact integral momentum balance
\begin{equation}
\epsilon\dot M
=
-\frac{1}{\mu}\left[Q^2-\frac{Q_u^2}{1+H}\right]
+\mu\left(H+\frac{H^2}{2}\right)
-\mu M,
\label{eq:sv_nonlinear_integral_balance}
\end{equation}
where $Q_u(t)\equiv\mathcal{Q}(0,t)$ denotes the upstream discharge. Since the balance is not closed in terms of $Q$ and $H$, we use the continuity relations below to combine the unresolved effects of $M$ and $Q_u$ into a single contribution.

For a fixed upstream stage $H$, the steady discharge $Q_s(H)$ is spatially uniform and satisfies
\begin{equation}
\frac{H}{\mu^2(1+H)}Q_s^2+Q_s=H+\frac{H^2}{2},
\label{eq:sv_steady_manifold_nonlinear}
\end{equation}
which defines the exact nonlinear steady-state family. Differentiating Eq.~\eqref{eq:sv_steady_manifold_nonlinear} with respect to $H$ gives
\begin{equation}
Q_s'(H)
=
\frac{\mu^2(1+H)^3-Q_s^2}
{(1+H)\left[\mu^2(1+H)+2H Q_s\right]}.
\label{eq:sv_steady_sensitivity_nonlinear}
\end{equation}
Substituting $Q=Q_s+\Delta Q$ and $M=Q_s+\Delta M$ into Eq.~\eqref{eq:sv_nonlinear_integral_balance} and using the steady balance in Eq.~\eqref{eq:sv_steady_manifold_nonlinear} yields
\begin{equation}
\epsilon Q_s'(H)\dot H+\epsilon\dot{\Delta M}
=
-\frac{2Q_s}{\mu}\Delta Q
-\frac{1}{\mu}(\Delta Q)^2
+\frac{Q_u^2-Q_s^2}{\mu(1+H)}
-\mu\Delta M.
\label{eq:sv_nonlinear_departure_balance}
\end{equation}
The unresolved quantities $\Delta M$ and $Q_u$ can be related to spatial moments of the free-surface displacement. Integration of the continuity equation gives
\begin{equation}
Q_u-Q=\epsilon\mu\dot{\mathcal{J}}_0,\qquad
M-Q=\epsilon\mu\dot{\mathcal{J}}_1,
\label{eq:sv_continuity_moments}
\end{equation}
where
\begin{equation}
\mathcal{J}_0(t)=\int_0^1\eta(x,t)\,dx,\qquad
\mathcal{J}_1(t)=\int_0^1x\,\eta(x,t)\,dx.
\label{eq:sv_spatial_moments}
\end{equation}

Substituting these relations into Eq.~\eqref{eq:sv_nonlinear_departure_balance} yields the exact form
\begin{equation}
\epsilon Q_s'(H)\dot H=-K_0(H)\Delta Q-N_0(H)(\Delta Q)^2-\epsilon\dot{\Delta Q}+\mathcal{U},
\label{eq:sv_nonlinear_departure_separated}
\end{equation}
where
\begin{equation}
K_0(H)=\mu+\frac{2H Q_s(H)}{\mu(1+H)},\qquad
N_0(H)=\frac{H}{\mu(1+H)},
\label{eq:sv_K0_N0}
\end{equation}
and
\begin{equation}
\mathcal{U}
=
\epsilon\left[
-\mu^2\dot{\mathcal{J}}_1
+\frac{2Q}{1+H}\dot{\mathcal{J}}_0
\right]
+\epsilon^2\mu\left[
-\ddot{\mathcal{J}}_1
+\frac{\dot{\mathcal{J}}_0^2}{1+H}
\right].
\label{eq:sv_unresolved_spatial}
\end{equation}
Equation~\eqref{eq:sv_nonlinear_departure_separated} separates the terms explicitly determined by the integral conservation law from the remaining unresolved spatial contribution, $\mathcal U$. In particular, the restoring terms involving $K_0(H)$ and $N_0(H)$, including the quadratic departure term $N_0(H)(\Delta Q)^2$, are retained in their exact conservation-derived form. The principal role of conservation is therefore to prescribe the known balance structure and restrict temporal closure to the unresolved spatial contribution $\mathcal U$.

Since the exact expression for $\mathcal U$ in Eq.~\eqref{eq:sv_unresolved_spatial} involves only time derivatives of spatial moments, we adopt a derivative-only temporal closure in departure coordinates. Following the general framework of Eq.~\eqref{eq:U_general_nondim}, we assume
\begin{equation}
\mathcal U\simeq
\sum_{k=1}^{p}\epsilon^k
\left[
C_{k,p}(H)\Delta Q^{(k)}
+D_{k,p}(H)H^{(k)}
\right].
\label{eq:SV_closure_p3_nonlinear}
\end{equation}
No zeroth-order departure term is introduced, so that the restoring terms determined by the integral conservation law are retained without modification. Substituting Eq.~\eqref{eq:SV_closure_p3_nonlinear} into Eq.~\eqref{eq:sv_nonlinear_departure_separated} and normalizing by $K_0(H)$ yields
\begin{equation}
\Delta Q+\frac{H(\Delta Q)^2}{\mu^2(1+H)+2H Q_s(H)}+\sum_{k=1}^{p}\epsilon^k A_{k,p}(H)\Delta Q^{(k)}=-\sum_{k=1}^{p}\epsilon^k B_{k,p}(H)H^{(k)},
\label{eq:sv_nonlinear_rom}
\end{equation}
where $A_{k,p}(H)$ and $B_{k,p}(H)$ are the normalized composite coefficients defined as follows:
\begin{equation}
\begin{aligned}
A_{1,p}&=\frac{1-C_{1,p}}{K_0}, \quad B_{1,p}=\frac{Q_s'-D_{1,p}}{K_0}, \\
A_{k,p}&=-\frac{C_{k,p}}{K_0},\quad B_{k,p}=-\frac{D_{k,p}}{K_0},\quad k\ge2.
\end{aligned}
\label{eq:nonlinear_closure_coeff}
\end{equation}

For stationary solutions under constant forcing, all temporal derivatives vanish, and Eq.~\eqref{eq:sv_nonlinear_rom} reduces exactly to the nonlinear steady-state balance in Eq.~\eqref{eq:sv_steady_manifold_nonlinear}, expressed in terms of the departure $\Delta Q$.  Although this closure is linear in the departure and its time derivatives, the ODE model remains nonlinear through the steady-state relation $Q_s(H)$, the state dependence of its dynamic coefficients, and the conservation-derived quadratic departure term. 

For the nonlinear benchmark, we adopt a third-order model without imposing the structured pole hierarchy used in the linear case. At a fixed operating state $\bar H$, we define local perturbations $\delta H=H-\bar H$ and $\delta Q=Q-Q_s(\bar H)$. Linearizing Eq.~\eqref{eq:sv_nonlinear_rom} about $\Delta Q=0$ eliminates the quadratic term and yields the normalized local transfer function
\begin{equation}
G^{\mathrm{loc}}_{Q}(\sigma;\overline H) \equiv \frac{\widehat{\delta Q}(\sigma)}{Q_s'(\overline H)\widehat{\delta H}(\sigma)}=\frac{1+\displaystyle\sum_{k=1}^{3}\left[A_{k,3}(\overline H)-\frac{B_{k,3}(\overline H)}{Q_s'(\overline H)}\right]\sigma^k}{
1+\displaystyle\sum_{k=1}^{3}A_{k,3}(\overline H)\sigma^k}.
\label{eq:sv_local_transfer_p3}
\end{equation}
This transfer function satisfies $G^{\mathrm{loc}}_Q(0;\bar H)=1$. The six coefficients $A{k,3}(\bar H)$ and $B_{k,3}(\bar H)$ are identified by matching six low-frequency moments of the semi-discrete PDE resolvent, as detailed in Appendix~A.2.

Because the downstream water depth is fixed, the flow must remain subcritical for the boundary conditions to be well posed. In subcritical flow, one characteristic enters the domain from the downstream boundary, allowing the water depth to be specified there. Along the steady-state family, the critical condition is first reached at the downstream boundary, where the fixed depth gives $\mathrm{Fr}=Q_s(H)/\mu$. Thus, subcritical steady states require $Q_s(H)<\mu$. Defining the corresponding critical upstream stage $H_c$ by $Q_s(H_c)=\mu$ and using Eq.~\eqref{eq:sv_steady_manifold_nonlinear} gives
\begin{equation}
H_c^3+3H_c^2-2\mu H_c-2\mu=0.
\label{eq:sv_critical_head}
\end{equation}
Accordingly, the operating range is restricted to $H<H_c(\mu)$ so that the steady solutions remain on the subcritical branch and the corresponding local linearizations retain the same well-posed boundary configuration.

For performance evaluation under finite-amplitude nonlinear conditions, we adopt the same exponential forcing and response-normalization forms as in Eq.~\eqref{eq:forcing_and_response}, with the Poiseuille variables $(F,q)$ replaced here by the upstream stage and downstream discharge $(H,Q)$. We consider acceleration from $H_0=0$ to $H_1=0.5$ and the corresponding deceleration from $H_1=0.5$ to $H_0=0$. Since $H$ is normalized by the reference depth, the forcing amplitude $\Delta H=0.5$ corresponds to a 50\% variation of that depth. Two dynamic regimes are examined, namely the inertia-dominated case $\mu=0.5$ and the friction-dominated case $\mu=4.0$, for forcing-timescale ratios $\epsilon=0.2$, $0.5$, and $1.0$. 

For both values of $\mu$, the operating range $0\le H\le0.5$ remains below the corresponding critical stage $H_c(\mu)$, so that all steady states lie on the subcritical branch. The full-order simulations are initialized from the numerically computed steady state at the initial value of $H$. For the third-order ODEs, we set $\Delta Q(0)=0$ and determine $\dot{\Delta Q}(0^+)$ and $\ddot{\Delta Q}(0^+)$ from the zero-state response of the corresponding locally linearized transfer function under the prescribed exponential forcing. For comparison, a mean-coefficient model is also considered, in which $A_{k,3}$ and $B_{k,3}$ are fixed at their averages over the operating range $H\in[0,0.5]$, while the nonlinear family of steady states and the conservation-imposed quadratic term remain state dependent. The normalized discharge response $Q_N(t)$ is defined in the same form as $q_N(t)$ in Eq.~\eqref{eq:forcing_and_response}, with $Q_0=Q_s(H_0)$ and $Q_1=Q_s(H_1)$.

Following the same offline construction and validation procedure as in Section~III.E, the Saint--Venant equations are spatially discretized on a uniform grid with $N_x=400$ cells using a cell-centered finite-volume scheme with Rusanov fluxes and characteristic boundary conditions. The choice of $N_x$ is based on a grid-sensitivity assessment. The resulting semi-discrete system of ODEs is integrated in time using MATLAB's ode15s solver with a relative tolerance of $10^{-9}$. The state-dependent third-order ODE is constructed from 25 anchor states over $H\in[0,0.5]$. At each anchor state, the corresponding steady solution is first obtained, after which the semi-discrete full-order system is locally linearized and the coefficients $A_{k,3}(H)$ and $B_{k,3}(H)$ are identified by matching the low-frequency moments of the local discharge transfer function. The identified ODE coefficients are interpolated over the operating range using piecewise cubic Hermite interpolation, while the steady-state quantities are evaluated analytically. The ODE is integrated with the ode15s solver using relative tolerances of $10^{-9}$. For the mean-coefficient model, the dynamic coefficients are fixed at their averages over $H\in[0,0.5]$, obtained by integrating the interpolated coefficient functions with respect to $H$ and dividing by the interval length.

\begin{figure}[t]
\centering
\includegraphics[width=1\linewidth]{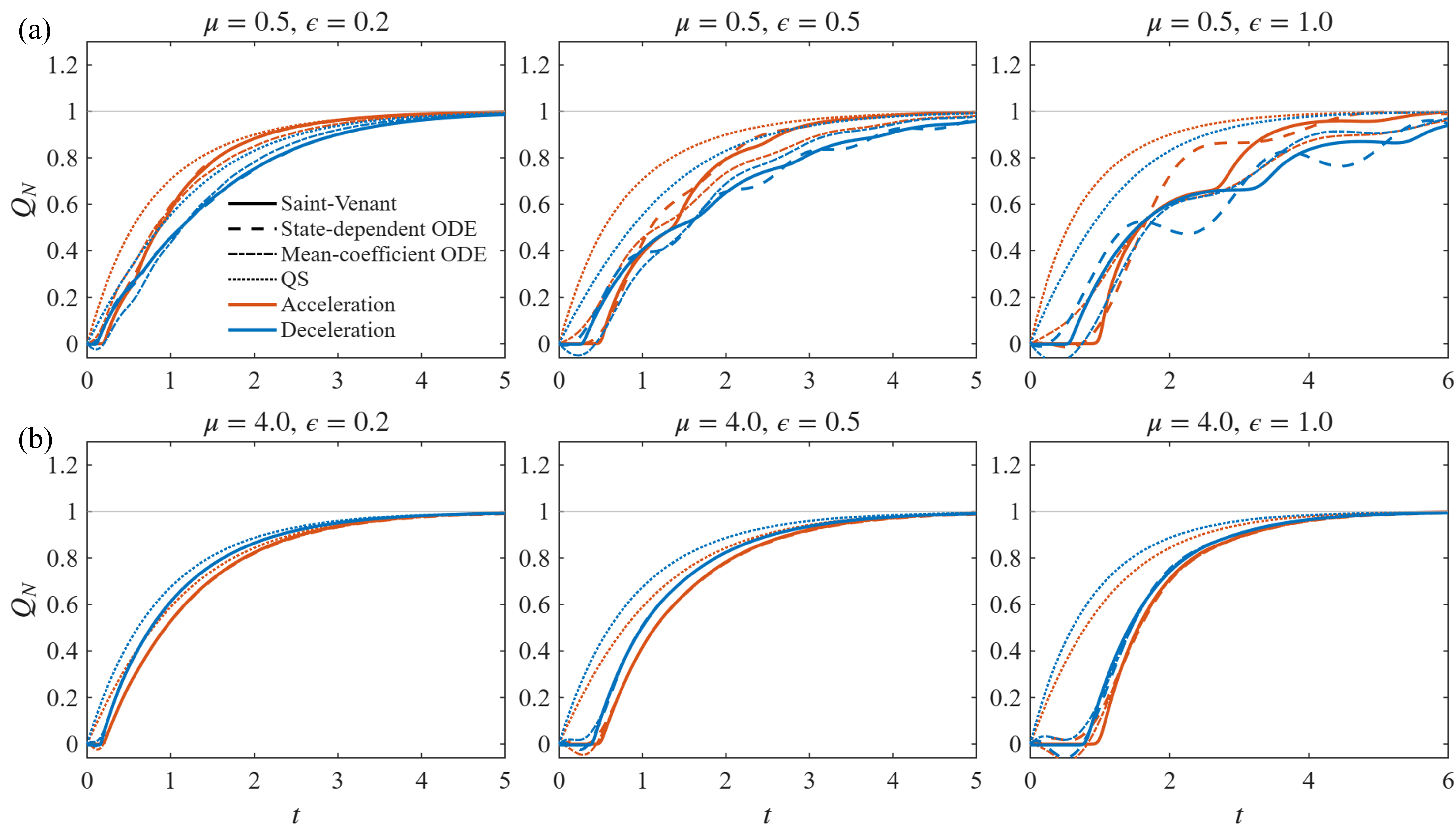}
\caption{Comparison of normalized transient discharge responses under exponential head forcing for acceleration from $H_0=0$ to $H_1=0.5$ (red) and deceleration from $H_1=0.5$ to $H_0=0$ (blue). Columns correspond to timescale ratios $\epsilon=0.2$, $0.5$, and $1.0$, and rows to (a) the inertia-dominated regime ($\mu=0.5$) and (b) the friction-dominated regime ($\mu=4.0$). Line styles denote the full nonlinear Saint--Venant model (solid), the state-dependent ODE ($p=3$, dashed), the mean-coefficient ODE ($p=3$, dash-dotted), and the quasi-steady approximation (dotted).}
\label{fig:6}
\end{figure}

Figure~\ref{fig:6} evaluates the performance of the ODE models against the full nonlinear Saint--Venant system across two distinct dynamic regimes ($\mu=0.5$ and $\mu=4.0$). For moderate forcing timescales ($\epsilon \le 0.5$), the proposed state-dependent third-order ODE closely matches the full PDE response in both regimes, capturing the finite-amplitude asymmetry between acceleration ($H_0=0 \to H_1=0.5$) and deceleration ($H_1=0.5 \to H_0=0$). The mean-coefficient ODE also reproduces the main features of the transient response. However, neither ODE model consistently outperforms the other, and their relative performance depends on the dynamical regime and forcing timescale.

In the friction-dominated regime ($\mu=4.0$), wave propagation is strongly attenuated, and the dynamic response becomes dominated by relaxation, similar to the behavior observed in the turbulent Poiseuille flow benchmark. Both the state-dependent and mean-coefficient models closely track the PDE results across all examined values of $\epsilon$, whereas the quasi-steady approximation responds more rapidly because it neglects transient storage and dynamic relaxation. The small difference between the two ODE models indicates that variations in the local dynamic coefficients play a minor role over the operating range considered, while transient storage and relaxation remain essential.

In the inertia/wave-dominated regime ($\mu=0.5$), discrepancies become more pronounced as $\epsilon$ increases, particularly in the phase of wave-induced oscillations associated with multiple boundary reflections. Under rapid forcing ($\epsilon=1.0$), both ODE models capture the overall transient response but exhibit phase mismatches in the higher-frequency oscillations. These discrepancies may be associated with nonlinear changes in wave-propagation times, as the characteristic speeds depend on the evolving spatial distributions of flow velocity and water depth, which are not fully represented by the locally linearized temporal closure. Additional small-amplitude calculations indicate that phase agreement improves toward the linear-response limit. This suggests a limitation of the present closure in representing finite-amplitude nonlinear wave propagation. Nevertheless, both reduced models reproduce the main transient features and retain the overall propagation and routing behavior even under faster forcing.
%==============================
\section{Discussion}
The two canonical benchmarks demonstrate that the appropriate structure and order of a temporal closure depend on the spectral properties of the underlying spatial dynamics and the frequency range excited by the forcing. The timescale ratio $\epsilon$ controls the relative importance of transient effects, while system-specific parameters determine their character. Plane Poiseuille flow is inherently relaxation-dominated because of viscous diffusion, whereas the dissipation parameter $\mu$ controls the relative importance of wave propagation and damping in the Saint--Venant system. Accordingly, low-order closures can capture the dominant response under strong damping, whereas higher-order closures are needed to reproduce wave-induced oscillations when propagation and reflection become significant.

In the inertia/wave-dominated Saint--Venant regime, weakly damped wave propagation and boundary reflections produce oscillatory dynamics involving complex-conjugate poles in addition to a real relaxation pole. A first-order closure cannot represent these oscillations. Although a second-order model can represent a complex-conjugate pole pair, the present Saint--Venant system also has a distinct real relaxation pole, making a third-order model the minimum order needed to retain both. Nevertheless, a single wave-mode pair cannot reproduce the full distributed wave spectrum. As forcing becomes more rapid, higher-frequency modes become increasingly important, and additional closure orders improve the representation of wave-induced oscillations.

The nonlinear Saint--Venant results further illustrate limitations of the present finite-order temporal closure with coefficients identified from local linear responses. The reduced model retains the exact nonlinear steady-discharge relation and conservation-derived restoring terms, while its temporal-closure coefficients are identified from locally linearized PDE responses. Under finite-amplitude forcing, the characteristic speeds depend on the evolving internal flow field, altering wave travel times and the interference of repeated reflections. Such effects are not fully characterized by linearization about the instantaneous steady state and may contribute to the phase discrepancies observed under weak friction and rapid forcing. The improved agreement at smaller forcing amplitudes supports this interpretation. Thus, matching local linear responses alone does not guarantee accurate prediction of finite-amplitude nonlinear wave propagation.

A further limitation of the low-frequency matching procedure adopted in these benchmarks concerns short-time and high-frequency behavior. The resulting finite-order ODE models may exhibit nonphysical initial responses, including nonzero initial flow-rate responses and impulse contributions to reconstructed wall-shear kernels in Poiseuille flow, and premature downstream responses before wave arrival in Saint--Venant flow. These artifacts reflect the limited high-frequency accuracy of the finite-order approximation of spatial memory rather than a deficiency in the conservation-derived balance structure. Incorporating high-frequency constraints into coefficient identification may reduce these artifacts, while reproducing an exact wave-arrival delay would require an explicit representation of propagation delay.

A central feature of the proposed framework is that the prescribed nonlinear steady-state relation is retained by construction, independently of the temporal closure order and the procedure used to determine its coefficients. This property follows from formulating the dynamic equations in terms of departures from the steady-state family and retaining the contributions explicitly determined by integral conservation laws. The identification of closure coefficients is a separate task. In the present benchmarks, these coefficients are obtained analytically or from local linearized resolvent moments without calibration against nonlinear transient trajectories. Other identification procedures can be adopted without changing the underlying modeling structure. 

Many engineering flow problems have well-established steady-state solutions or relations, yet the quasi-steady approximation is insufficient to describe their transient responses. The present framework offers a systematic way to extend such steady-state descriptions to unsteady conditions while preserving the prescribed steady-state relations and the conservation-derived balance structure. It is particularly applicable when the dynamics evolve around a smooth family of steady states and the unresolved spatial effects can be adequately represented by a finite-order temporal closure. Strongly nonequilibrium flows, hydraulic jumps, self-sustained instabilities, and bifurcations remain outside the scope of this study. Extensions to higher-dimensional systems may require more elaborate representations of the steady-state family and suitable strategies for identifying temporal closure coefficients.

\section{Conclusion}
This study developed a modeling framework that extends the quasi-steady approximation by representing unsteady fluid responses as dynamic departures from a family of steady solutions. Integral conservation laws guide the construction of the departure equations, while finite-order temporal closures represent unresolved spatial dynamics. A key feature of the framework is that the prescribed nonlinear steady-state relations are recovered exactly under constant forcing. The canonical validations show that the appropriate dynamic structure depends on both the forcing timescale and the underlying spatial memory: low-order relaxation dynamics can be sufficient when higher spatial modes contribute weakly to the selected bulk response, whereas wave propagation and reflection require richer temporal structures to represent oscillatory memory.

The nonlinear viscous-diffusion and shallow-water benchmarks further demonstrate that state-dependent dynamic coefficients can be identified without calibration against unsteady trajectories. However, the nonlinear Saint--Venant results reveal phase discrepancies under rapid, finite-amplitude forcing, suggesting that the present finite-order temporal closure, with coefficients identified from local linear responses, does not fully capture nonlinear changes in wave propagation and reflection. Addressing these effects may require richer closure representations and alternative coefficient-identification strategies.

\begin{acknowledgments}
During the preparation of this manuscript, the author used a generative AI tool (ChatGPT, GPT-5.6 and GPT-6) for language improvement and manuscript consistency checks. After using this tool, the author reviewed and edited the content as needed and takes full responsibility for the content of this work.
\end{acknowledgments}

\section*{Data Availability}
The numerical data and MATLAB codes supporting the findings of this study are available from the corresponding author upon reasonable request.

\appendix
\section{Coefficient Identification via Local Resolvent Moment Matching}
\label{app:resolvent_matching}
%==========================================================================================
This appendix details the offline identification of state-dependent reduced-model coefficients using low-frequency resolvent moment matching along the family of steady states. Adopting the general framework notation of Section~II ($\lambda$ for forcing, $Y$ for bulk output, and $Y_s(\lambda)$ for the steady-state family), this procedure is based on classical moment-matching theory (e.g., \cite{Antoulas2005}), applying local resolvent expansions along the family of steady states.

Consider a fixed operating point $\bar{\lambda}$ on the family of steady states. Spatially discretizing the governing equations and linearizing the resulting semi-discrete system about this steady state yields the semi-discrete linear perturbation system
\begin{equation}
\epsilon \frac{\mathrm{d}\boldsymbol{X}}{\mathrm{d}t} = \boldsymbol{J}(\bar{\lambda}) \boldsymbol{X} + \boldsymbol{b}(\bar{\lambda}) \, \delta \lambda(t), \qquad \delta Y(t) = \boldsymbol{c}(\bar{\lambda})^T \boldsymbol{X}(t),
\label{eq:app_semidiscrete}
\end{equation}
where $\boldsymbol{X}(t)$ is the discretized perturbation state vector (representing velocity-field perturbations $\delta \boldsymbol{u}$ for viscous dynamics in Benchmark~I, or coupled stage-discharge perturbations $[\delta \boldsymbol{\eta}, \delta \boldsymbol{\mathcal{Q}}]^T$ for wave propagation dynamics in Benchmark~II). The matrix $\boldsymbol{J}(\bar{\lambda})$ is the Jacobian of the semi-discrete governing equations evaluated at the steady state, while $\boldsymbol{b}(\bar{\lambda})$ and $\boldsymbol{c}(\bar{\lambda})^T$ are the discrete input distribution and output observation vectors, respectively.

In practice, the steady-state solution is sampled at a finite set of anchor operating points of $\lambda$ and the following identification procedure is applied independently at each anchor state. At each anchor state, $\boldsymbol{J}(\bar{\lambda})$, $\boldsymbol{b}(\bar{\lambda})$ and $\boldsymbol{c}(\bar{\lambda})^T$ are obtained by linearizing the semi-discrete governing equations about the computed steady solution.

Taking the Laplace transform of Eq.~\eqref{eq:app_semidiscrete}, we normalize the local semi-discrete transfer function by its zero-frequency gain:
\begin{equation}
G^{\mathrm{loc}}_{Y}(\sigma;\bar{\lambda}) \equiv \frac{\boldsymbol{c}(\bar{\lambda})^T[\sigma\boldsymbol{I}-\boldsymbol{J}(\bar{\lambda})]^{-1}\boldsymbol{b}(\bar{\lambda})}{h_0(\bar{\lambda})},\qquad h_0(\bar{\lambda})=-\boldsymbol{c}(\bar{\lambda})^T\boldsymbol{J}(\bar{\lambda})^{-1}\boldsymbol{b}(\bar{\lambda}).
\label{eq:app_local_transfer}
\end{equation}
Here, $h_0(\bar{\lambda})$ is the exact steady gain of the semi-discrete linearization, ensuring $G^{\mathrm{loc}}_{Y}(0;\bar{\lambda})=1$ at finite grid resolution. It approaches the continuous steady-state sensitivity $Y_s'(\bar{\lambda})$ ($q_s'(\bar F)$ for Benchmark~I or $Q_s'(\bar H)$ for Benchmark~II) under grid refinement. The continuous sensitivity is retained in the reduced-order model and in the conversion of matched moments to forcing-side coefficients. Expanding the resolvent in powers of $\sigma$ about $\sigma=0$ gives the low-frequency moment series \citep{Antoulas2005}
\begin{equation}
G^{\mathrm{loc}}_{Y}(\sigma;\bar{\lambda})=1+\sum_{k=1}^{\infty}m_k(\bar{\lambda})\sigma^k,\qquad m_k(\bar{\lambda})=-\frac{\boldsymbol{c}(\bar{\lambda})^T\boldsymbol{J}(\bar{\lambda})^{-(k+1)}\boldsymbol{b}(\bar{\lambda})}{h_0(\bar{\lambda})}.
\label{eq:app_moments}
\end{equation}

For the stable operating points considered here, $\boldsymbol{J}(\bar{\lambda})$ is nonsingular, allowing the moments $m_k(\bar{\lambda})$ to be computed algebraically through repeated linear solves without unsteady time integration. Then, the state-dependent reduced-model coefficients are obtained by matching the coefficients of Eq.~\eqref{eq:app_moments} with those of the reduced transfer function for each benchmark problem. Finally, the coefficients are interpolated along the family of steady states. Further details of the determination process for each problem are given as follows.

\subsection{Identification for Benchmark~I: viscous-diffusion dynamics}
For the turbulent viscous-diffusion benchmark ($\lambda=F$, $Y=q$), the frozen-state local flow-rate transfer function is given by Eq.~\eqref{eq:turbulent_local_transfer}. Expanding Eq.~\eqref{eq:turbulent_local_transfer} in powers of $\epsilon s$ and matching coefficients with Eq.~\eqref{eq:app_moments} through $\mathcal{O}(\sigma^2)$ uniquely determines the local flow-rate closure coefficients as
\begin{equation}
\begin{aligned}
C_0(\bar F)&=\frac{m_1(\bar F)}{m_2(\bar F)},\\
D_1(\bar F)&=q_s'(\bar F)\left[1-C_0(\bar F)m_1(\bar F)\right].
\end{aligned}
\label{eq:app_bm1_coefficients}
\end{equation}

\subsection{Identification for Benchmark~II: wave-propagation dynamics}
For the nonlinear Saint--Venant benchmark ($\lambda=H$, $Y=Q$), the normalized local discharge transfer function derived from the dynamic-departure model is given by Eq.~\eqref{eq:sv_local_transfer_p3}. Cross-multiplying Eq.~\eqref{eq:sv_local_transfer_p3} with the low-frequency expansion in Eq.~\eqref{eq:app_moments} yields the following matching relation through sixth order:
\begin{equation}
\begin{aligned}
\left(1+\sum_{k=1}^{3}A_{k,3}(\bar H)\sigma^k\right)\left(1+\sum_{k=1}^{\infty}m_k(\bar H)\sigma^k\right)=1+\sum_{k=1}^{3}\left[A_{k,3}(\bar H)-\frac{B_{k,3}(\bar H)}{Q_s'(\bar H)}\right]\sigma^k+\mathcal O(\sigma^7).
\end{aligned}
\label{eq:app_bm2_identity}
\end{equation}

Because the right-hand side of Eq.~\eqref{eq:app_bm2_identity} contains no terms beyond third order in $\sigma$, equating the coefficients of $\sigma^4$, $\sigma^5$, and $\sigma^6$ determines the denominator coefficients $A_{k,3}(\bar H)$ through the $3\times3$ moment system involving the higher resolvent moments:
\begin{equation}
\begin{bmatrix}
m_3(\bar H)&m_2(\bar H)&m_1(\bar H)\\
m_4(\bar H)&m_3(\bar H)&m_2(\bar H)\\
m_5(\bar H)&m_4(\bar H)&m_3(\bar H)
\end{bmatrix}
\begin{bmatrix}A_{1,3}(\bar H)\\[0.8ex]A_{2,3}(\bar H)\\[0.8ex]A_{3,3}(\bar H)\end{bmatrix}
=-\begin{bmatrix}m_4(\bar H)\\[0.8ex]m_5(\bar H)\\[0.8ex]m_6(\bar H)\end{bmatrix}.
\label{eq:app_bm2_A_system}
\end{equation}
Once $A_{k,3}(\bar H)$ are obtained by solving Eq.~\eqref{eq:app_bm2_A_system}, equating the coefficients of $\sigma^1$, $\sigma^2$, and $\sigma^3$ in Eq.~\eqref{eq:app_bm2_identity} determines the corresponding forcing-side composite coefficients $B_{k,3}(\bar H)$:
\begin{equation}
\begin{aligned}
B_{1,3}(\bar H)&=-Q_s'(\bar H)\,m_1(\bar H),\\
B_{2,3}(\bar H)&=-Q_s'(\bar H)\left[A_{1,3}(\bar H)m_1(\bar H)+m_2(\bar H)\right],\\
B_{3,3}(\bar H)&=-Q_s'(\bar H)\left[A_{2,3}(\bar H)m_1(\bar H)+A_{1,3}(\bar H)m_2(\bar H)+m_3(\bar H)\right].
\end{aligned}
\label{eq:app_bm2_coefficients}
\end{equation}

Thus, both the denominator coefficients $A_{k,3}(\bar H)$ and the forcing-side composite coefficients $B_{k,3}(\bar H)$ are determined algebraically from the resolvent moments $m_1(\bar H)$ through $m_6(\bar H)$, without requiring unsteady snapshot simulations. The corresponding closure coefficients $C_{k,3}(\bar H)$ and $D_{k,3}(\bar H)$ follow from the normalized departure balance in the main text.

\nocite{*}

\bibliography{apssamp}% Produces the bibliography via BibTeX.

\end{document}